\documentclass[aps,prb,twocolumn]{revtex4}
\usepackage{amssymb,amstext,amsmath}
\usepackage{graphicx}
\usepackage{float}
\usepackage[pdftex]{hyperref}
\newcommand{\He}{$^{3}$He}

\renewcommand{\Gauge}{\mathsf{U}(1)}
\newcommand{\SO}{\mathsf{SO}}
\newcommand{\Civ}{\mathsf{C}_\mathrm{\infty v}}
\newcommand{\Dih}{\mathsf{D}_\mathrm{\infty h}}
\newcommand{\Dth}{\mathsf{D}_\mathrm{2 h}}
\newcommand{\ESP}{\mathsf{SO}(2)_\mathrm{S_d} \times \mathsf{Z}_2^\mathrm{spin}}
\newcommand{\Ab}[2]{A_{#1 #2}^{*}}
\newcommand{\Ag}[3]{A_{#1 #2 , #3}}
\newcommand{\nm}{\,\mbox{nm}}
\newcommand{\mbar}{\,\mbox{bar}}
\newcommand{\angstrom}{\textup{\AA}}
\newcommand{\Pz}{$\text{P}_\mathbf{z}$}
\newcommand{\Bi}{$\text{B}_\mathbf{SO(2)}$}
\newcommand{\Ai}{$\text{A}_\mathbf{SO(2)}$}
\newcommand{\At}{$\text{A}_\mathbf{C_2}$}
\newcommand{\wpre}{\frac{\gamma^{2}}{\chi_{N}}g_{D}}
\newcommand{\wpreB}{\frac{\gamma^{2}}{\chi_{B}}g_{D}}
\newcommand{\savg}[1]{\left\langle #1 \right\rangle}
\newcommand{\Drr}{\left\langle \Delta_r^2 \right\rangle}
\newcommand{\Dtt}{\left\langle \Delta_\phi^2 \right\rangle}
\newcommand{\Dzz}{\left\langle \Delta_z^2 \right\rangle}
\newcommand{\Dxx}{\left\langle \Delta_x^2 \right\rangle}
\newcommand{\Dyy}{\left\langle \Delta_y^2 \right\rangle}
\newcommand{\Drz}{\left\langle \Delta_r \Delta_z \right\rangle}
\newcommand{\Dtz}{\left\langle \Delta_\phi \Delta_z \right\rangle}
\newcommand{\Drt}{\left\langle \Delta_r \Delta_\phi \right\rangle}
\newcommand{\TAB}{T_{\mbox{\tiny AB}}}
\newcommand{\pPCP}{p_{\mbox{\tiny PCP}}}
\newcommand{\betawc}[1]{\beta_{#1}^{\text{wc}}}
\newcommand{\betasc}[1]{\beta_{#1}^{\text{sc}}}
\newcommand{\sn}[2]{$#1 \hspace{-0.3em}\times \hspace{-0.4em}10^{#2}$}
\newcommand{\scb}[1]{\Delta\beta^{\text{sc}}_{#1}}
\def\nicefrac#1#2{\genfrac{}{}{}{1}{#1}{#2}}
\begin{document}
\title{Superfluid phases of \He\ in nano-scale channels}
\author{J. J. Wiman}
\author{J. A. Sauls}
\affiliation{Department of Physics and Astronomy, Northwestern University, Evanston, Illinois 60208}
\date{\today}
\begin{abstract}
Confinement of superfluid $^3$He on length scales comparable to the radial size of the p-wave Cooper pairs
can greatly alter the phase diagram by stabilizing broken symmetry phases not observed in bulk \He. 
We consider superfluid $^3$He confined within long cylindrical channels of radius $100\mbox{ nm}$,
and report new theoretical predictions for the equilibrium superfluid phases under strong confinement. 
The results are based on the strong-coupling formulation of Ginzburg-Landau theory with precise numerical 
minimization of the free energy functional to identify the equilibrium phases and their regions of stability.
We introduce an extension of the standard GL strong-coupling theory that accurately accounts for the 
phase diagram at high pressures, including the tri-crital point and $T_{AB}(p)$ line defining the region of
stability for the bulk A-phase.
We also introduce tuneable boundary conditions that allow us to explore boundary scattering ranging from 
maximal to minimal pairbreaking, and report results for the phase diagram as a function of pressure, 
temperature, and boundary conditions.
Four stable phases are found: a polar phase stable in the vicinity of $T_c$, a strongly anisotropic, cylindrical 
analog of the bulk B phase stable at sufficiently low temperatures, and two chiral A-like phases with 
distinctly different orbital symmetry, one of which spontaneously breaks rotation symmetry about the axis of 
the cylindrical channel. The relative stability of these phases depends sensitively on pressure and the degree of 
pairbreaking by boundary scattering.
The broken symmetries exhibited by these phases give rise to distinct signatures in transverse NMR resonance
spectroscopy. We present theoretical results for the transverse NMR frequency shifts as functions of temperature,
the {\sl rf} pulse tipping angle and the static NMR field orientation.
\end{abstract}
\maketitle

\vspace*{-3mm}
\section{Introduction}
\vspace*{-3mm}

Superfluid \He{} is a spin-triplet, p-wave Fermi superfluid, where not only is $\Gauge{}_N$ symmetry 
spontaneously broken but also spin and orbital rotation symmetries $\SO{}(3)_S \times \SO{}(3)_L$. There 
are a myriad of ways to break these symmetries, leading to many potential superfluid phases. In bulk \He{}, in 
the absence of a magnetic field, only two stable phases are observed: the A phase and the B 
phase. However, other phases may be stabilized by introducing symmetry breaking terms, such as a magnetic field, 
impurities, or boundaries, which couple to the spin and orbital degrees of freedom of the Cooper pairs. 
In particular, confining surfaces suppress Cooper pairs with relative momentum normal to the surface, 
which leads to a long-range orienting effect on the orbital order parameter.\cite{amb75} When confined within 
distances comparable to the Cooper pair coherence length, $\xi_0 \approx 160-770 \,\angstrom{}$ depending on pressure, 
the influence of the confining surfaces can stabilize phases 
much different than those of bulk superfluid \He{}.

Advances in nanoscale fabrication techniques,\cite{lev13} as well as the production of porous materials 
with interesting structure on the coherence length scale,\cite{pol12,ask12} have made studies of the 
effects of strong confinement on broken symmetry phases of topological quantum materials feasible, and have brought 
a surge of research on the effects of confinement on superfluid \He{}.\cite{chu09,nom14,miz11,miz12,wu13}
One of the simplest confining geometries is the pore, 
a long, small radius cylinder. The pore has long been of theoretical interest due to the number of 
different A-phase textures that might be stabilized,\cite{mak78} as well the effects of radial confinement 
on the superfluid phase diagram.\cite{fet88,tak87,li88} Nuclear magnetic resonance (NMR) experiments in 
$2 \,\mathrm{\mu m}$ diameter pores have 
observed A-like textures,\cite{gou78,sau78} but have had difficulty definitively identifying the 
textures present.\cite{bru79} New fabrication techniques for porous membranes\cite{mas95} have made available 
pores with diameters below $1 \,\mathrm{\mu m}$, which, coupled with an array of new experimental 
techniques,\cite{gon13,zhe14,lev13,roj15} open new windows into superfluid \He{} under strong confinement.

In this paper we consider an infinitely long cylindrical pore of radius $R=100\nm{}$, and we study the 
equilibrium phases in Ginzburg-Landau (GL) theory and identify their signatures in 
nonlinear NMR spectroscopy. By incorporating 
pressure dependent strong-coupling corrections to the GL material coefficients, and a tuneable
pairbreaking boundary condition, we obtain phase diagrams as functions of temperature, pressure, and 
surface condition. Finally, we derive expressions for the transverse NMR frequency shifts of the  
equilibrium phases of \He{} confined in the pore as functions of {\sl rf} pulse driven tipping angle, and 
show how they vary with order parameter symmetry and orientation of the static magnetic field.

\vspace*{-3mm}
\section{Ginzburg-Landau Theory}
\vspace*{-3mm}

We use Ginzburg-Landau theory calculations of the superfluid \He{} order parameter and free energy to 
determine the stable phases present in the pore. The order parameter for superfluid \He{}, given by the 
manifold of spin-triplet, p-wave BCS pairing states, may be represented by the $2 \times 2$ gap matrix,
\begin{equation}
\hat{\Delta}(\hat{p}) = \sum_{\alpha i} A_{\alpha i}\, (i \sigma_\alpha \sigma_y)\, \hat{p}_i \,,
\end{equation}
which depends on the direction of the relative momentum $\hat{p}$ of the Cooper pairs, and is parameterized 
by the $3 \times 3$ complex matrix order parameter $A$. The matrix $A$ transforms as a vector under spin rotations, 
and separately as a vector under orbital rotations. In cylindrical coordinates $A_{\alpha i}$ can be represented as
\begin{equation}
A = 
\begin{pmatrix} 
A_{rr} & A_{r \phi} & A_{rz} \\
A_{\phi r} & A_{\phi \phi} & A_{\phi z} \\
A_{zr} & A_{z\phi} & A_{zz} 
\end{pmatrix}
\,,
\end{equation}
where we have chosen aligned spin and orbital coordinate axes.

The presence of boundaries reduces the possible residual orbital symmetries of the superfluid phases to 
be elements of the point group of the confining cylindrical geometry. However, this reduction in 
symmetry is due to interactions atomically close to the boundary surface; away from the surface, the 
\He{} particle-particle interactions are still invariant under the maximal symmetry group of bulk \He{}.
Thus, the Ginzburg-Landau free energy functional is given by the invariants of the bulk \He{} symmetry group,
\begin{equation}
\mathsf{G}_{\mathrm{bulk}} = \Gauge \times \SO(3)_\mathrm{S} \times \SO(3)_\mathrm{L} 
\times \mathsf{P} \times \mathsf{T} \,, 
\end{equation}
which is the product of global gauge rotations, spin rotations, orbital rotations, space inversion, and 
time-reversal, respectively. The resulting free energy functional is
\begin{equation}
\Omega[A] = \int_{V} d^3r\,\left( f_{\mathrm{bulk}}[A] + f_{\mathrm{grad}}[A] \right) 
\,.
\end{equation}
The terms $f_\mathrm{bulk}$ and $f_{\mathrm{grad}}$ are given by
\begin{widetext}
\begin{align} 
f_\mathrm{bulk}[A] &=  
\alpha(T) Tr\left(A A^{\dagger}\right) 
+\beta_{1} \left|Tr(A A^{T})\right|^{2}
+\beta_{2} \left[Tr(A A^{\dagger})\right]^{2}  
\nonumber \\
& \qquad
+\beta_{3}\, Tr\left[A A^{T} (A A^{T})^{*}\right] 
+\beta_{4}\, Tr\left[(A A^{\dagger})^{2}\right]
+\beta_{5}\, Tr\left[A A^{\dagger} (A A^{\dagger})^{*}\right] \,,
\end{align}
\begin{align}
f_\mathrm{grad}[A] &=
K_{1} A_{\alpha j , k}^{*} A_{\alpha j , k}
+K_{2} A_{\alpha j , j}^{*} A_{\alpha k , k}
+K_{3} A_{\alpha j , k}^{*} A_{\alpha k , j}
\nonumber \\	
&\quad +\frac{2}{r}\, \mathrm{Re}\left\lbrace
K_1 \left( \Ab{r}{j} \Ag{\phi}{j}{j}
-\Ab{\phi}{j}\Ag{r}{j}{\phi}
+\Ab{i}{r}\Ag{i}{\phi}{\phi}
-\Ab{i}{\phi}\Ag{i}{r}{\phi} \right) \right.
\nonumber \\
&\qquad\qquad \left.
+K_2 \left( \Ab{r}{\phi} \Ag{\phi}{j}{j} 
-\Ab{\phi}{\phi} \Ag{r}{j}{j} 
+\Ab{i}{r} \Ag{i}{j}{j} \right) \right.
\nonumber \\
&\qquad\qquad \left.	 
+K_3 \left( \Ab{r}{j}\Ag{\phi}{\phi}{j}
-\Ab{\phi}{j}\Ag{r}{\phi}{j}
+\Ab{i}{r}\Ag{i}{\phi}{\phi}
-\Ab{i}{\phi}\Ag{i}{\phi}{r} \right) \right\rbrace
\nonumber \\	
&\quad +\frac{1}{r^2}\, \left\lbrace
K_1 \left\lbrack \Ab{r}{j}A_{rj}+\Ab{\phi}{j}A_{\phi j}
+\Ab{i}{r}A_{ir}+\Ab{i}{\phi}A_{i\phi}
+4\mathrm{Re}(\Ab{r}{\phi}A_{r\phi}-\Ab{r}{r}A_{\phi \phi}) \right\rbrack
\right.
\nonumber \\
&\qquad\qquad \left.
+(K_2+K_3)\left\lbrack |A_{r\phi}|^2+|A_{\phi \phi}|^2 +\Ab{i}{r}A_{ir}
+ 2\mathrm{Re}(\Ab{r}{\phi}A_{\phi r}-\Ab{r}{r}A_{\phi \phi})	\right\rbrack
\right\rbrace
\,,
\end{align}
\end{widetext}
where $A^{\dag}$ ($A^{T}$) is the adjoint (transpose) of $A$, and
\begin{equation}
A_{\alpha i , j} \equiv \left\lbrace
\frac{\partial A_{\alpha i}}{\partial r}\,,\; \frac{1}{r}\frac{\partial A_{\alpha i}}{\partial \phi}\,,\;
\frac{\partial A_{\alpha i}}{\partial z} \right\rbrace_j \,.
\end{equation}
The term $f_\mathrm{bulk}$ holds for any orthogonal coordinate system, whereas $f_{\mathrm{grad}}$ 
is coordinate specific and given in the form derived by Buchholtz and Fetter.\cite{buc77}
In the weak-coupling BCS limit the material parameters,
\begin{eqnarray}
\alpha^{\text{wc}}(T) 
&=& 
\frac{1}{3}N(0)(T/T_{c}-1)\,,
\\
2\beta_{1}^{\text{wc}} 
&=& -\beta_{2}^{\text{wc}} = -\beta_{3}^{\text{wc}} = -\beta_{4}^{\text{wc}} = \beta_{5}^{\text{wc}}
\,,
\label{eq-beta-wc}
\\
\beta_{1}^{\text{wc}}
&=&
-\frac{N(0)}{(\pi k_{\text{B}}T_{c})^2}\left\{\frac{1}{30}\left[\frac{7}{8}\zeta(3)\right]\right\}
\,,
\\
K_{1}^{\text{wc}} 
&=&
K_{2}^{\text{wc}} = K_{3}^{\text{wc}} = \frac{7 \zeta(3)}{60} N(0)\, \xi_{0}^2
\,,
\end{eqnarray}
are determined by the normal-state (single-spin) density of states at the Fermi energy, $N(0)$, the bulk 
superfluid transition temperature, $T_c$, and the Fermi velocity, $v_f$. 
Note that
$\xi_{0} = \hbar v_{f} / 2\pi k_{B} T_{c}$ is the Cooper pair correlation 
length, which varies from $\xi_{0}\simeq 770\,\angstrom$ at $p=0\,\mbox{bar}$ to 
$\xi_{0}\simeq 160\,\angstrom$ at $p = 34\,\mbox{bar}$. The equilibrium order parameter is 
obtained from minimization of the free energy functional 
by solving the Euler-Lagrange equations obtained from the functional gradient 
$\delta \Omega[A]/\delta A^\dagger=0$.

\vspace*{-3mm}
\subsection{Strong-coupling Corrections}
\vspace*{-3mm}

\begin{figure}[t]
	\begin{center}
		\includegraphics[width=2.7in]{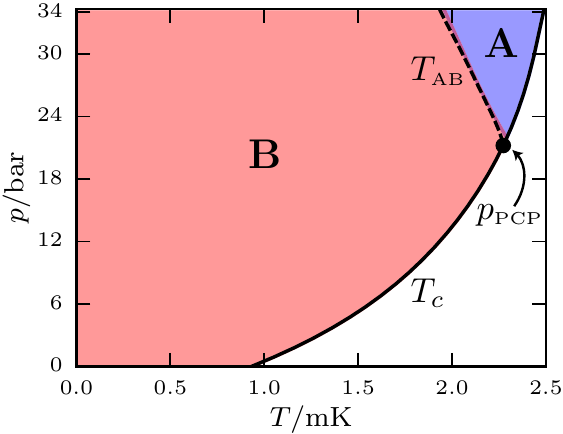}
		\caption{The bulk phase diagram showing the regions of stability of the A (blue) 
		         and B (red) phases using the experimental $\beta$s with linear $T$ scaling. 
			 The dashed line is the experimental A-B transition line terminating at the 
			 experimental PCP point.}
		\label{bulk-phase-diagram}
	\end{center}
\end{figure}

The weak-coupling GL material parameters are derived from the leading order contribution to the full 
Luttinger-Ward free energy functional as an expansion in the small parameter $T/T_F$, 
where $T_F=E_F/k_{B}\approx 1\,\mbox{K}$ is the Fermi temperature. 
In particular, $\Omega^\text{wc} \sim \left(T_c/T_F\right)^2E_N$, where $E_N$ is the 
ground-state energy of the normal Fermi liquid. The next-to-leading corrections to the weak-coupling GL functional 
enter as corrections to the fourth-order weak-coupling material coefficients. These corrections are of order 
$\Delta\betasc{i}\approx\betawc{i}(T/T_F)\langle w_{i}|A|^2\rangle$, 
where $\langle w_{i}|A|^2\rangle$ is a weighted average of the square of the scattering amplitude for binary 
collisions between quasiparticles on the Fermi surface.\cite{rai76}
At high pressures, scattering due to ferromagnetic spin fluctuations largely compensates the small 
parameter $T/T_F$, resulting in substantial strong-coupling corrections.\cite{sau81b}

While the $\Delta\betasc{i}$'s may be calculated theoretically through a model of the 
quasiparticle scattering amplitude\cite{sau81b}, the most current determinations come from comparison
with experiment.\cite{cho07} 
In the main analysis presented here we use the set of $\{\beta_i\}$ reported by Choi et al.\cite{cho07} 
These $\beta$-parameters reproduce the heat capacity jumps for the 
A and B transitions, which is essential when considering the energetics of A and B-like phases. 
In particular, the A phase correctly appears as a stable phase above the polycritical 
point $\pPCP{}=21.22\mbar{}$; however, in fourth-order GL theory it is the only stable 
phase at all temperatures above the PCP, i.e. the standard fourth-order GL theory fails to account 
for the A-B transition line, $\TAB{}(p)$. The missing transition line is traced to the omission of the 
temperature dependence of the fourth-order $\beta$ parameters in the neighborhood of a tri-critical point. 
In particular, the tri-critical point is defined by the intersection of the second-order transition line 
given by $\alpha(T_c,p) = 0$, and the first-order boundary line separating the A- and B-phases given by 
$\Delta\beta_{AB}(\TAB{},p)\equiv\beta_{A}-\beta_{B}=0$.
Note that $\beta_{A}\equiv\beta_{245}$ and $\beta_{B}\equiv\beta_{12}+\nicefrac{1}{3}\beta_{345}$ where we use the 
stanadard notation, $\beta_{ijk\ldots}=\beta_i+\beta_j+\beta_k+\ldots$.\cite{rai76} 
At the PCP we have $\TAB{}(\pPCP{})=T_c(\pPCP{})$. But, for $p>\pPCP{}$ the lines separate and we must retain 
both the temperature and pressure dependences of $\Delta\beta_{AB}(T,p)$ to account for 
$\TAB{}(p)$ in the vicinity of $\pPCP{}$. 
This is achieved with remarkable success by making a single correction to the standard treatment of
strong-coupling corrections within GL theory.
Near $T_c$ the leading-order strong-coupling corrections to the weak-coupling $\beta$ parameters scale as 
$\Delta\betasc{i}\sim(T/T_F)|\betawc{1}|$, where the linear scaling with $T/T_F$ originates from the limited 
phase space for binary collisions of quasiparticles at low temperatures. 
Resolving the degeneracy between the A- and B-phases near $\pPCP{}$ is achieved by retaining the linear $T$ 
dependence of the strong-coupling corrections to the $\beta$ parameters.
Thus, we separate the $\beta$ parameters determined at $p$ and $T_c(p)$ into the weak- and strong-coupling parts 
using Eq. \ref{eq-beta-wc}, and then scale the strong-coupling corrections, $\Delta\betasc{i}$, determined 
at $T_c(p)$ in Ref. \onlinecite{cho07} and listed in Tables \ref{table:3He-Material-Parameters} and 
\ref{table:3He-Material-Parameters_Fit-Choi} of the Appendix,
\begin{align}
\label{eq-betas_sc-scaling}
\beta_i(T,p) 
&= \betawc{i}(p,T_c(p))+\frac{T}{T_c}\Delta\betasc{i}(p) 
\,,
\\ 
\mbox{with}\quad
\Delta\betasc{i}(p)
&= \beta_i(p,T_c(p))-\betawc{i}(p,T_c(p)) 
\,.
\end{align}

The resulting bulk phase diagram predicted by these GL parameters accounts remarkably well for the 
experimental A-B transition line, $\TAB{}(p)$, as shown in Fig. \ref{bulk-phase-diagram}, as well as the 
heat capacity jumps and the PCP along $T_c(p)$. This result for the bulk phase diagram gives us confidence 
in our predictions for the equilibrium phases of confined \He{} based on strong-coupling GL theory. 
The main analysis and predictions for inhomogeneous phases of superfluid \He{} reported here are based on the 
strong-coupling material parameters from Ref. \onlinecite{cho07} combined with the known pressure-dependent 
material parameters, $v_f$, $T_c$, and $\xi_0$ as listed in Table \ref{table:3He-Material-Parameters} in the 
Appendix, and the temperature scaling in Eq. \ref{eq-betas_sc-scaling} that accounts for the relative reduction 
of strong-coupling effects below $T_c$. 

\vspace*{-3mm}
\subsection{Sauls-Serene $\beta$ parameters}
\vspace*{-3mm}

The individual $\Delta\beta^{\text{sc}}_{i}$ parameters reported by Choi et al.\cite{cho07}
differ from those calculated from leading order strong-coupling theory, or  
those obtained from the analysis of different experiments, even though the different sets predict 
the same bulk phase diagram.

As a test of the sensitivity of our GL predictions for new phases in confined geometries
to the details of the model for the strong-coupling GL $\beta$ parameters we also calculated the
phase diagram based on the $\{\beta_i\}$ predicted by the leading order strong-coupling theory.\cite{rai76,sau81b}
The theoretical values for the strong-coupling $\beta$ parameters are determined by angular averages of the 
normal-state quasiparticle scattering rate. The analysis of Sauls and Serene is based on a quasiparticle
scattering amplitude that accounts for the effective mass, the ferromagnetic enhancement of the spin 
susceptibility and the normal-state transport coefficients.\cite{sau81b} 
The Sauls-Serene $\beta$-parameters, summarized in Tables \ref{table:3He-Material-Parameters-SS81}
and \ref{table:3He-Material-Parameters_Fit-SS81}, reproduce the relative stability of the bulk 
A and B phases, albeit with an elevated polycrital pressure of $p_{\text{PCP}}\simeq 28\,\mbox{bar}$.

The results for the phase diagram with these two different sets of $\Delta\beta_{i}^{\text{sc}}$, discussed in 
Sec. \ref{sec-Phase_Diagram}, give robust predictions for the relative stabilty of new inhomogeneous phases of \He\ 
confined in cylindrical pores.

\vspace*{-3mm}
\subsection{Boundary Conditions}
\vspace*{-3mm}

For planar surfaces there are two limiting boundary conditions applicable within GL theory: 
maximal pairbreaking, resulting from retro-reflection of quasiparticles,\cite{sau11} 
and minimal pairbreaking, resulting from specular reflection.\cite{amb75} 
If we use cartesian coordinates and take the surface to lie along the $x-y$ plane with \He{} occupying $z>0$, 
then maximal pairbreaking is defined by the condition
\begin{align}
A_{\alpha i}\big|_{z=0} &= 0\,\,\forall i\in\{x,y,z\}\,,
\end{align}
while minimal pairbreaking is defined by the conditions
\begin{align}
A_{\alpha z}\big|_{z=0} &= 0\,, \nonumber \\
\nabla_z A_{\alpha x}\big|_{z=0} &= \nabla_z A_{\alpha y}\big|_{z=0} = 0\,.
\end{align}

In a cylindrical pore, additional care needs to be given to the boundary conditions due to the 
presence of curvature on scales comparable to the coherence length. While the boundary condition for 
maximal pairbreaking is not modified, the curved surface of the pore modifies the minimal pairbreaking 
boundary condition for the azimuthal orbital components of the order parameter, $A_{\alpha\phi}$. Fetter 
and Buchholtz proposed a minimal pairbreaking boundary condition in GL theory based on the Euler-Lagrange 
boundary term of the GL equations with a cylindrical surface,\cite{buc77} 
\begin{eqnarray}
&&
\frac{\partial A_{\alpha z}}{\partial r}\Big\vert_{r=R} = 0
\,,
\quad 
A_{\alpha r}\vert_{r=R} = 0
\,,
\\
&&
\frac{\partial A_{\alpha \phi}}{\partial r}\Big\vert_{r=R} = \frac{1}{R}A_{\alpha \phi}\vert_{r=R}
\,.
\end{eqnarray}
We introduce an extension of these boundary conditions which interpolates between the two extremes of minimal and 
maximal pairbreaking. The extension is based on Ambegaokar, de Gennes, and Rainer's (AdGR) analysis\cite{amb75} of 
the effects of diffuse scattering by an atomically rough surface on the transverse components of the 
p-wave orbital order parameter. AdGR showed that diffuse scattering leads to a boundary condition in which 
the components that are transverse to the average normal direction of the surface are finite, but extrapolate 
linearly to zero past the boundary at a distance $b_T=0.54\xi_0$. This idea can be turned into a more general 
boundary condition for GL theory in a cylindrical geometry as 
\begin{eqnarray}
A_{\alpha r}\vert_{r=R} &=& 0
\,, 
\nonumber\\
\frac{\partial A_{\alpha z}}{\partial r}\Big\vert_{r=R} 
&=& 
-\frac{1}{b_T}A_{\alpha z}\vert_{r=R}
\,, 
\nonumber\\
\frac{\partial A_{\alpha \phi}}{\partial r}\Big\vert_{r=R} 
&=& 
\left(\frac{1}{R}-\frac{1}{b_T}\right)A_{\alpha \phi}\vert_{r=R}
\,.
\label{GL-boundary_conditions}
\end{eqnarray}
where $b_T^{\prime}\equiv b_T/\xi_0$ can be treated as a parameter that varies between the maximal pairbreaking 
($b_T^{\prime}\rightarrow 0$) and minimal pairbreaking ($b_T^{\prime}\rightarrow\infty$) limits. This 
generalized ``AdGR'' boundary condition provides a useful extension of the typical Ginzburg-Landau boundary conditions.

\vspace*{-3mm}
\section{Superfluid Phases}
\vspace*{-3mm}

The pore geometry reduces the maximal symmetry group for confined \He{} to 
\begin{equation}
\mathsf{G} = \Gauge_{\mathrm{N}} \times \SO(3)_\mathrm{S} \times \Dih \times \mathsf{T}\,
\end{equation}
where $\Dih{}$ is the point group of the pore and is obtained from the point group of the circle, 
$\Civ{} = \SO(2) \times \lbrace \mathrm{e},\, \pi_{z x} \rbrace$, by 
$\Dih{} = \Civ{} \times \lbrace \mathrm{e},\, \pi_{x y} \rbrace$,
where $\pi_{x y}$ is a reflection through the $x-y$ plane. By numerically minimizing the GL free energy with respect 
to all order parameter components we identify four equilibrium superfluid phases for the $200\nm{}$ pore. In these 
calculations we assume the phases are translationally invariant along the $z$ axis.\cite{Note1}

\vspace*{-3mm}
\subsection{Polar (\Pz{}) Phase}
\vspace*{-3mm}

\begin{figure}[t]
\begin{center}
\includegraphics[width=2.4in]{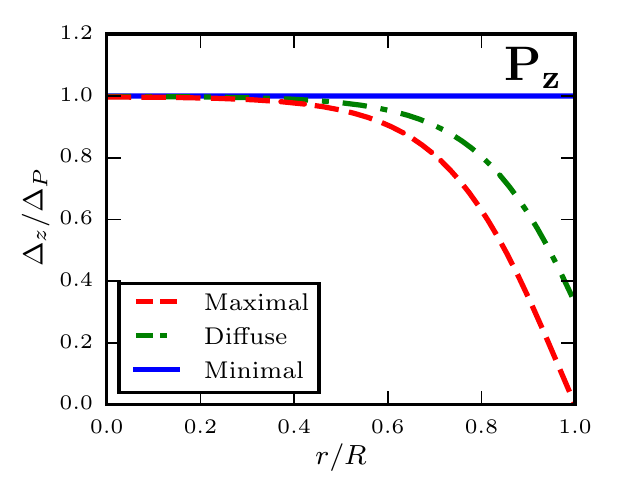}
\caption{Order parameter amplitudes for the \Pz{} phase as a function of $r$ at $p=26\,\mbox{bar}$ and 
$T=0.5\,T_c$ for retro-reflection (maximal pair-breaking), diffuse ($b_T^{\prime}=0.54$), and specular (minimal 
pair-breaking) boundary conditions. Values are scaled by the unconfined polar phase amplitude 
$\Delta_P^2=|\alpha(T)|/2\beta_{12345}$. For minimal pairbreaking boundary conditions the \Pz{} phase is 
spatially homogeneous within the pore.}
\label{op_p}
\end{center}
\end{figure}

Radial confinement in a cylindrical pore leads to the stability of the one-dimensional polar (\Pz{}) 
phase below $T_{c_1} \leq T_c$, where $T_{c_1}$ is the transition temperature from the normal state. 
The \Pz{} phase is a time-reversal invariant equal-spin pairing (ESP) phase with an order parameter of form
\begin{equation}
\label{eq:pz}
A_{\alpha i} = \Delta_z(r)\, \hat{d}_\alpha\, \hat{z}_i
\,,
\end{equation}
with radial profile shown in Fig. \ref{op_p}. 
The \Pz{} order parameter becomes spatially homogeneous with $T_{c_1}\rightarrow T_c$ in the limit of specular 
scattering, and will be the first superfluid phase upon cooling from the normal state, except for the exceptional 
case of perfect specular reflection and perfect cylindrical cross-section (see Sec. \ref{sec-Ai}).
The residual symmetry group of the \Pz{} phase is $\mathsf{H}=\ESP\times\Dih^\mathrm{L,\pi}\times\mathsf{T}$, 
where $\Dih^\mathrm{L,\pi}\equiv\Civ{}\times\lbrace\mathrm{e},\, e^{i\pi}\pi_{x y}\rbrace$. Thus, the \Pz{} phase 
breaks spin rotational symmetry but retains the full orbital point group, $\Dih$, by combining it with an element 
of the gauge group. 
Since the radius $R=100\nm{}$ of the pore is much less than the dipole coherence length, 
$\xi_{\text{D}}\approx 10-20\,\mu\mbox{m}$, the spin quantization axis, $\hat{d}$, for the ESP state 
is to high accuracy uniform in space. All transitions to and from the \Pz{} phase that we find are second order.

\vspace*{-3mm}
\subsection{\Bi{} Phase}
\vspace*{-3mm}

\begin{figure}[t]
\begin{center}
\includegraphics[width=1\linewidth]{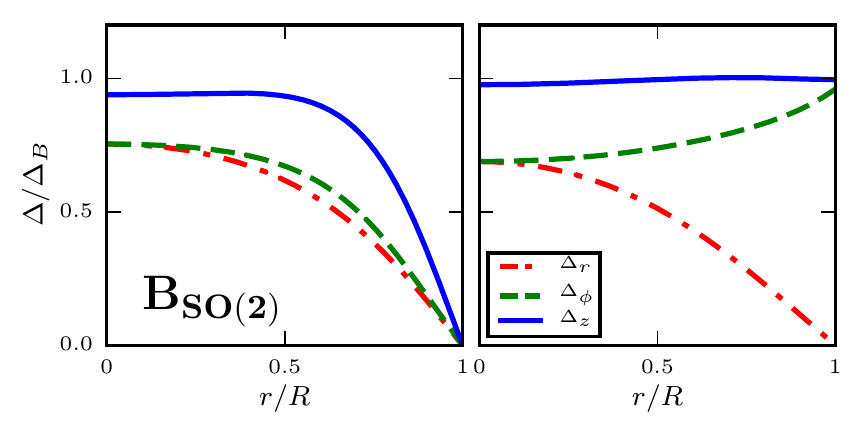}
\caption{Order parameter amplitudes for the \Bi{} phase as a function of $r$ at $p=26\,\mbox{bar}$ and 
$T=0.5\,T_c$. The left figure depicts maximal pairbreaking ($b_T'=0$) while the right shows minimal 
pairbreaking ($b_T'=\infty$). Values are scaled by the bulk B phase order parameter, 
$\Delta_B^2=|\alpha(T)|/6(\beta_{12}+1/3\beta_{345})$.}
\label{op_b}
\end{center}
\end{figure}

The \Bi{} phase is the analogue to the bulk B phase for the cylindrical pore geometry, and is stabilized at low 
temperatures and preferentially favored by strong pairbreaking on the boundary. The residual symmetry of the \Bi{} 
phase is $\mathsf{H} = \Dih^\mathrm{L+S} \times \mathsf{T}$,
joint spin and orbital $\Dih{}$ transformations combined with time-reversal.
The order parameter is represented as
\begin{equation}\label{eq:bi}
A_{\alpha i} = \Delta_r(r) \hat{r}_\alpha \hat{r}_i + \Delta_\phi(r) \hat{\phi}_\alpha \hat{\phi}_i
+ \Delta_z(r) \hat{z}_\alpha \hat{z}_i \,,
\end{equation}
with the radial profiles shown in Fig. \ref{op_b}.

\vspace*{-3mm}
\subsection{\Ai{} Phase}\label{sec-Ai}
\vspace*{-3mm}

\begin{figure}[t]
\begin{center}
\includegraphics[width=1\linewidth]{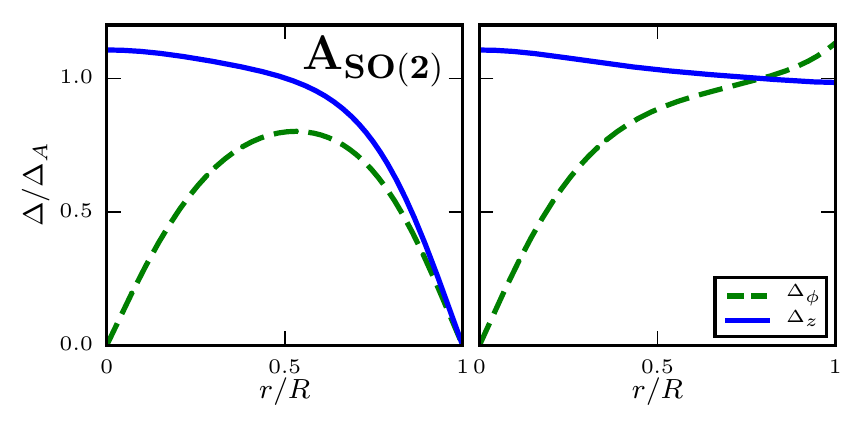}
\caption{Order parameter amplitudes for the \Ai{} phase as a function of $r$ at $p=26\,\mbox{bar}$ and 
$T=0.5\,T_c$. The left figure depicts maximal pairbreaking while the right shows minimal pairbreaking. 
Values are scaled by the bulk A phase order parameter, $\Delta_A^2=|\alpha(T)|/4\beta_{245}$.}
\label{op_ci}
\end{center}
\end{figure}

\begin{figure}[t]
\begin{center}
\includegraphics[width=2.2in]{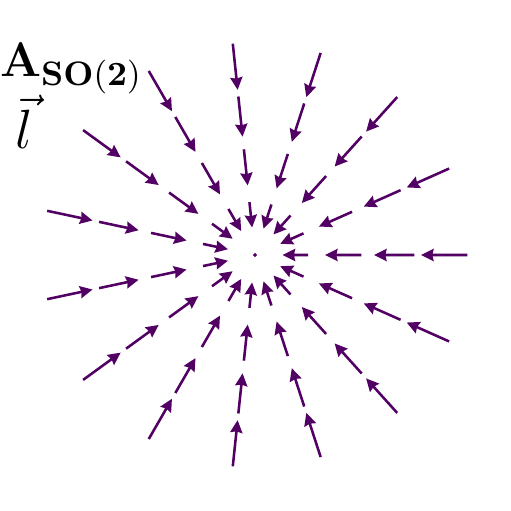}
\caption{Chiral axis $\hat{l}$ for the \Ai{} phase at $p=26\,\mbox{bar}$, $T=0.5\,T_c$, and minimal 
pairbreaking boundaries.
Arrow lengths are scaled by $|\Delta_\phi|$.}
\label{op_ci_chiral}
\end{center}
\end{figure}

In addition to the \Pz{} and \Bi{} phases, we find two stable chiral A-like phases. The higher symmetry \Ai{} 
phase, reminiscent of the ``radial disgyration texture'' of bulk \He-A, is favored by weak pair-breaking on 
the boundary. The residual symmetry group of the \Ai{} phase is 
$\mathsf{H} = \ESP \times \Dih^\mathrm{L,T}$, where 
$\Dih^\mathrm{L,T} \equiv \SO(2) \times \lbrace \mathrm{e},\, \mathrm{t} \pi_{z x} \rbrace
\times \lbrace \mathrm{e},\,e^{i\pi} \mathrm{t} \pi_{x y} \rbrace$ and $\mathrm{t}$ is time reversal. 
The order parameter takes the form
\begin{equation}\label{eq:ai}
A_{\alpha i} = \hat{d}_\alpha \lbrack \Delta_z(r) \hat{z}_i + i\,\Delta_\phi(r) \hat{\phi}_i \rbrack
\end{equation} 
which is transverse to the pore boundary with radial profiles shown in Fig. \ref{op_ci}. 
The chiral vector,
\begin{equation}\label{chiral_axis}
\vec{l} = -\Delta_z(r)\Delta_\phi(r)\,\hat{r}
\,,
\end{equation}
shown in Fig. \ref{op_ci_chiral}, is radial except at 
the origin where $\vec{l}$ vanishes; the gradient terms in the GL functional require $\Delta_\phi=0$, 
yielding a polar order parameter at the core of the cylindrical pore.
The resulting chiral field is analogous to a radial disgyration - a topological line defect of bulk \He-A.  
This form for $\vec{l}$ results in a zero average of $\vec{l}(r)$ over the cylindrical pore, which leads
to distinctly different NMR frequency shift for the \Ai{} phase as compared with a chiral 
state with a non-vanishing average chiral axis, $\langle\vec{l}(\vec{r})\rangle \ne 0$, as discussed in 
Sec. \ref{sec-NMR}. 
For the specular boundary condition proposed in Ref. \onlinecite{buc77} (Eq. \ref{GL-boundary_conditions} 
with $b_T'\rightarrow\infty$), 
the \Ai{} phase entirely supplants the \Pz{} phase, and onsets at the bulk transition temperature $T_c$,
despite being spatially inhomogeneous.


\vspace*{-3mm}
\subsection{\At{} Phase}
\vspace*{-3mm}

\begin{figure}[t]
\begin{center}
\includegraphics[width=1\linewidth]{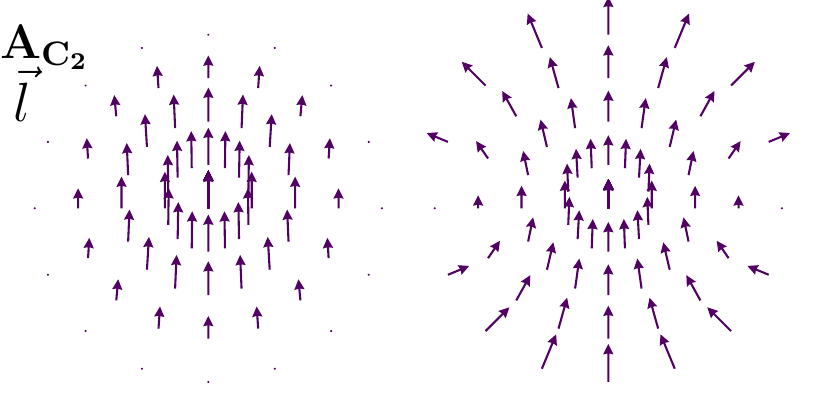}
\caption{Chiral axis $\hat{l}$ for the \At{} phase at $p=26\,\mbox{bar}$ and $T=0.5\,T_c$. (Left) Maximal 
pairbreaking boundaries result in a nearly uniform $\hat{l}$ direction. (Right) Minimal pairbreaking, 
on the other hand, gives the characteristic ``Pan Am'' configuration.
Arrow lengths are scaled by $(\Delta_r^2+\Delta_\phi^2)^{1/2}$.}
\label{op_ct_chiral}
\end{center}
\end{figure}

A lower symmetry A-like phase, denoted as \At{}, is an inhomogenous version of the the circular disgyration, 
or Pan Am texture.\cite{mak78,gou78}
This phase spontaneously breaks continuous $\SO(2)_{L}$ symmetry of the cylinder and, 
unlike the \Ai{} phase, has a finite value for the spatially averaged chiral axis, $\langle\vec{l}\rangle\ne 0$,
that may point in any direction in the $x-y$ plane.
For convenience we take $\langle\vec{l}\rangle \parallel \hat{y}$, with an order parameter of the form
\begin{equation}\label{eq:at}
A_{\alpha i} = \hat{d}_\alpha 
\lbrack 
\Delta_z(r,\phi) \hat{z}_i 
+
i\,\Delta_r(r,\phi) \hat{r}_i 
+ 
i\,\Delta_\phi(r,\phi) \hat{\phi}_i 
\rbrack 
\,.
\end{equation} 
The residual symmetry group is then 
$\mathsf{H} = \ESP \times \Dth^\mathrm{L,T}$, where 
$\Dth^\mathrm{L,T} \equiv 
\lbrace \mathrm{e},\, \mathrm{t} c_2,\, \pi_{zx},\, \mathrm{t} \pi_{zy} \rbrace
\times \lbrace \mathrm{e},\, e^{i\pi} \mathrm{t} \pi_{x y} \rbrace$.
The \At{} phase has a pair of disgyrations on the boundary along an axis perpendicular to $\hat{z}$ and 
$\langle\vec{l}\rangle$, as can be seen for the case of minimal pairbreaking in Fig. \ref{op_ct_chiral}.
The \At{} phase is energetically favorable relative to the \Ai{} phase for strong pairbreaking on the boundary. 
In this case the boundary effectively ``censors'' the energy cost of the \At{} disgyrations. The suppression 
of the disgyrations is evident in Fig. \ref{op_ct_chiral}.

The \At{} phase is the only phase we find with broken axial symmetry, and thus 
explicit $\phi$ dependence. It is convenient to 
expand its amplitudes in terms of sines and cosines that respect symmetry,
\begin{align}
A_{\alpha i} 
&= \hat{d}_\alpha 
\sum_{j=0}^{\infty} 
\left\lbrace 
\vphantom{\sum}
i\,\Delta_{r,j}(r)\cos[(2j+1)\phi]\,\hat{r}_i 
\right. 
\nonumber\\
&\quad 
\left. 
-i\,\Delta_{\phi,j}(r)\sin[(2j+1)\phi]\, \hat{\phi}_i 
+ 
\Delta_{z,j}(r)\cos(2j\phi) \,\hat{z}_i 
\right\rbrace 
\,.
\end{align}
Numerical solutions to the GL equations converge rapidly as a function of the number of azimuthal
harmonics, which greatly simplifies the numerical minimization compared to allowing for an 
arbitrary $\phi$ dependence.

\begin{figure}[t]
\begin{center}
\includegraphics[width=1.0\linewidth]{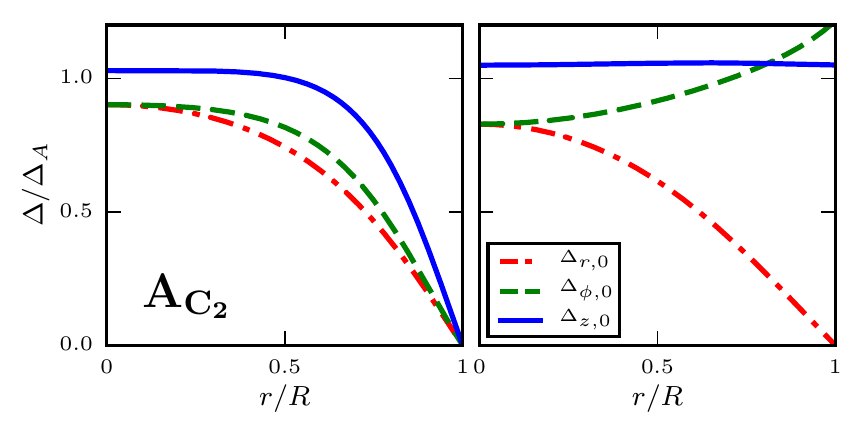}
\caption{Dominant order parameter amplitudes for the \At{} phase at $p=26\,\mbox{bar}$ and $T=0.5\,T_c$. 
The left figure depicts maximal pairbreaking while the right shows minimal pairbreaking. Values are 
scaled by the bulk A phase amplitude $\Delta_A^2=|\alpha(T)|/4\beta_{245}$.}
\label{op_ct}
\end{center}
\end{figure}

\vspace*{-3mm}
\section{Phase Diagram}\label{sec-Phase_Diagram}
\vspace*{-3mm}

\begin{figure}[t]
\begin{center}
\includegraphics[width=3.4in]{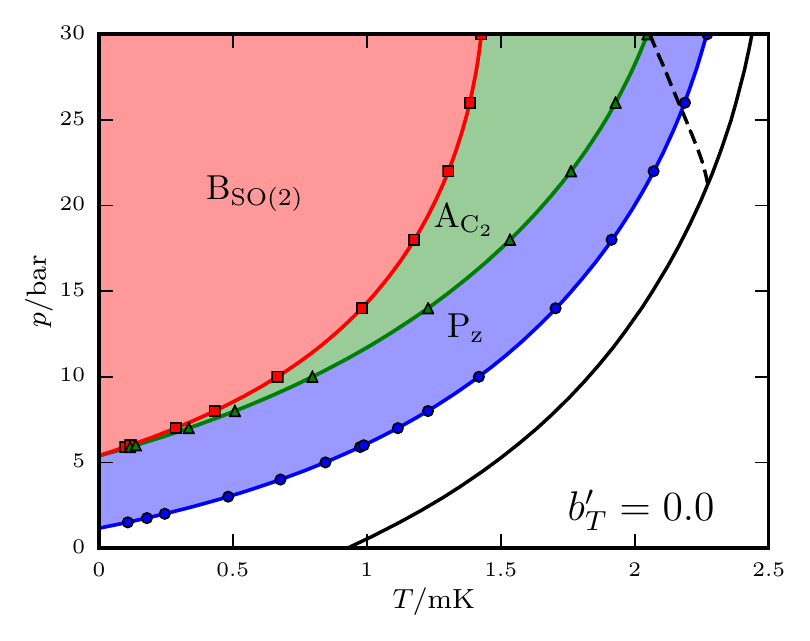}
\includegraphics[width=3.4in]{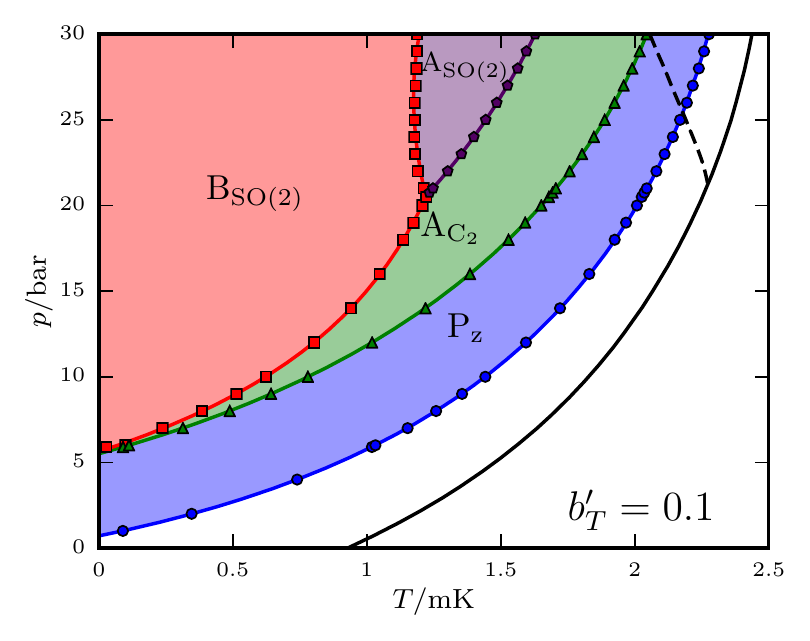}
\caption{Phase diagrams for $b_T'=0$ (maximal pairbreaking) and $b_T'=0.1$. 
The \Ai{} phase does not appear at all for maximal pairbreaking; as pairbreaking at the boundary is 
relaxed it is stabilized at high pressure and displaces the \Bi{} and \At{} phases.}
\label{phase_diagram_max}
\end{center}
\end{figure}

\begin{figure}[t]
\begin{center}
\includegraphics[width=3.4in]{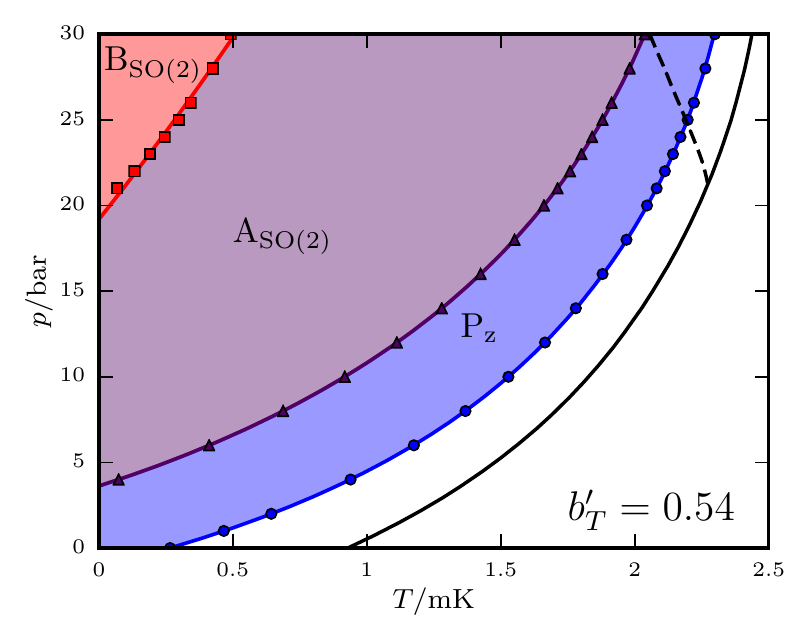}
\includegraphics[width=3.4in]{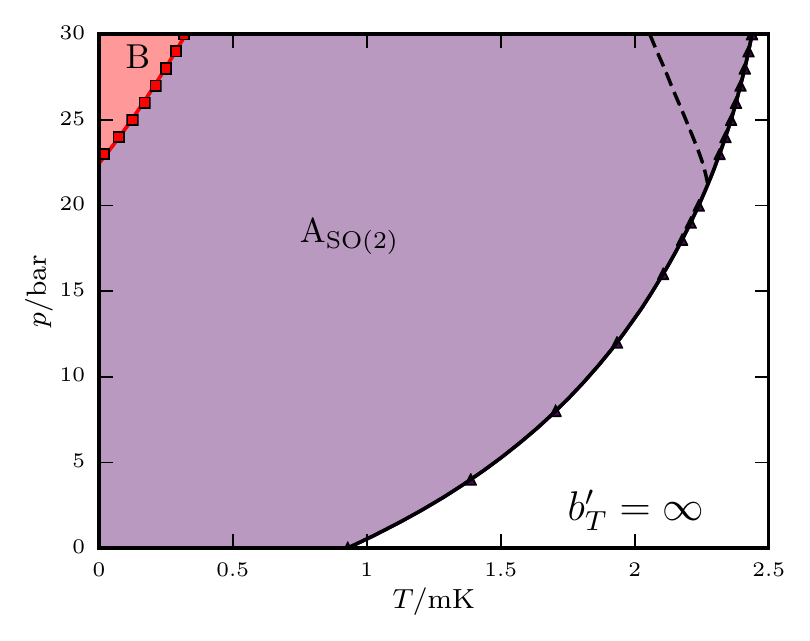}
\caption{Phase diagrams for $b_T'=0.54$ and $b_T'=\infty$ (minimal pairbreaking). 
As pairbreaking decreases, the \At{} phase is suppressed completely and the stable range of the \Bi{} 
phase is decreased significantly. For minimal pairbreaking the \Ai{} phase onsets at $T=T_c$ with the 
\Pz{} phase absent.}
\label{phase_diagram_min}
\end{center}
\end{figure}

The phase diagram for superfluid \He{} confined within a pore is strongly dependent upon the 
boundary conditions. We first fix $R=100\nm{}$ and consider the phase diagram for four different 
values of $b_T'$, ranging from minimal to maximal pairbreaking as shown in Figures \ref{phase_diagram_max} 
and \ref{phase_diagram_min}. For strong pairbreaking (Fig. \ref{phase_diagram_max}) the phase 
diagram is dominated by the \Bi{}, \At{}, and \Pz{} phases.
In this regime, our phase diagram differs from previous calculations\cite{li88,fet88} due to the 
appearance of the \At{} phase, which for strong pairbreaking has a lower free energy than that of 
the \Ai{} phase. 
As pairbreaking decreases on the boundary, the \Ai{} phase appears at high pressure, with a tri-critical point
separating the \Bi, \Ai{} and \At{} phases.
The \Ai{} phase occupies 
most of the superfluid phase diagram for minimal pairbreaking boundaries. It must be noted, however, 
that any deviation from the perfect specular condition $b_T'=\infty$ will suppress $A_{\alpha \phi}$ at 
the boundary near $T_c$, and thus the \Pz{} phase should always be expected to be the highest 
temperature superfluid phase observed experimentally.

\begin{figure}[t]
\begin{center}
\includegraphics[width=3.4in]{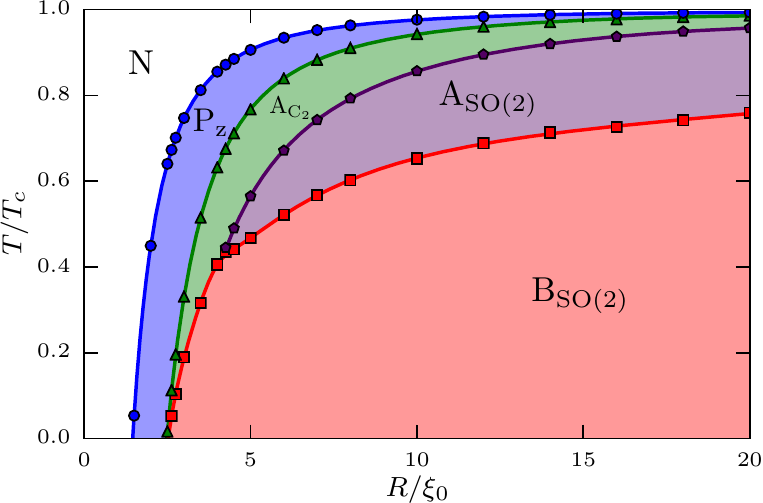}
\caption{Temperature-confinement phase diagram for fixed $b_T'=0.1$, $p=26\,\mbox{bar}$
         and the $\beta$ parameters of Ref. \onlinecite{cho07}.}
\label{phase_diagram_td}
\end{center}
\end{figure}

We also consider the influence of the pore radius, $R$, on stability of the various phases. 
Fig. \ref{phase_diagram_td} shows the phase diagram of stable phases in a cylindrical channel 
as a function of the pore radius relative to the coherence length, $R/\xi_0$.
For a range of sufficiently small $R/\xi_0$ only the \Pz{} phase is stable; 
the \Bi{}, \At{}, and \Ai{} phases enter the diagram with increasing $R$. 
The \Ai{} phase is favored over the \At{} phase for large radii; however, the relative stability of these
two chiral phases is sensitive to boundary scattering, i.e. $b_T'$, as shown in Figs. 
\ref{phase_diagram_max} and \ref{phase_diagram_min}. 
For larger radii of order the dipole coherence length, $R\approx\xi_D\approx 10\,\mu\mbox{m}$, the spin 
quantization axis, $\hat{d}$, for the \At{} and \Ai{} phases is no longer constrained to be spatially uniform,
and for $R\gg\xi_D$ these phases become ``dipole-locked'' with $\hat{d}\parallel\hat{l}$.\cite{buc77}

We also tested the robustness of our predictions for the phase diagram against a different set of strong-coupling 
$\beta$ parameters, specifically the Sauls-Serene set of $\Delta\beta_{i}^{\text{sc}}$ calculated on the basis 
of leading order strong-coupling theory\cite{rai76} based on a quasiparticle scattering amplitude that 
accounts for both the normal-state effective mass, ferromagnetic enhancement of the spin susceptibility and 
transport coefficients.\cite{sau81b} These $\beta$ parameters account for the relative stability of the bulk
A- and B-phases, but have distinctly different predictions for the pressure dependences of the strong-coupling 
corrections: $\Delta\beta_{i}^{\text{sc}}$.
The key result is that the structure of the phase diagram is unchanged with a different set of strong-coupling
$\beta$ parameters, i.e. the relative stability of the \Pz{}, \Bi{}, \At{}, and \Ai{} phases is unchanged between
the two sets of strong-coupling $\beta$ parameters. This is shown in Fig. \ref{phase_diagram_td}.


\vspace*{-3mm}
\section{NMR Signatures}\label{sec-NMR}
\vspace*{-3mm}

The superfluid phases obtained for the narrow pore neglect the nuclear magnetic dipole energy. This is
an execellent approximation since the  nuclear dipole-dipole interaction energy for dipoles separated by 
the mean interatomic spacing $a$ is very small compared to the pairing energy scale, 
$\nicefrac{1}{a^3}\,(\gamma\hbar/2)^2\approx 10^{-4}\,\text{mK}\ll T_c\approx \text{mK}$.
Nevertheless, the dipole energy gives rise to two important effects - (i) it partially resolves relative spin-
orbital degeneracy of the equilibrium states, and (ii) it generates a dynamical torque from the Cooper pairs 
acting on the total spin when the latter is driven out of equilibrium. The dipolar torque leads to NMR frequency 
shifts that are characteristic signatures of the broken symmetry phases. In the following we report results for the
nonlinear NMR frequency shifts that are ``fingerprints'' of the \Pz, \Ai, \At{}, and \Bi{} phases. Our analysis 
is based on a spatial and temporal averaging of the Leggett equations for the nonlinear spin dynamics of 
superfluid \He{}.

The dipolar interaction breaks relative spin-orbit rotation symmetry, 
thus reducing the maximal rotational symmetry from 
$\SO(3)_\mathrm{S} \times \SO(3)_\mathrm{L}$ to $\SO(3)_\mathrm{L+S}$. 
This is reflected by additional terms in the GL free energy functional,  
$\Delta \Omega_{D} = \int_{V} d^3r\,f_D[A]$, where
\begin{equation}
f_{D} = g_D \left( |Tr{A}|^2  + Tr{AA^*} \right)
\label{eq-dipole_energy}
\end{equation}
is the mean pairing contribution to the nuclear dipolar interaction energy, with 
$g_{D}\sim (N(0)\gamma\hbar/2)^2 > 0$. 
A convenient expression for $g_D$ is $g_{\text{D}} = \nicefrac{\chi}{2\gamma^2}\Omega_A^2/\Delta_A^2$, 
where the A-phase susceptibility, $\chi=\chi_N$, is equal to the normal-state spin susceptibility, 
$\Delta_A$ is the bulk A-phase order parameter, and
$\Omega_A$ is the corresponding longitudinal resonance frequency.  
The dipole energy is a weak perturbation that resolves (partially) the relative spin- and orbital degeneracy of the 
zero-field phases of the cylindrical pore.
In particular, for the ESP states of the form 
$A_{\alpha i}=\hat{d}_{\alpha}\,\Delta_{i}(\vec{r})$
the dipolar energy is given by $f_{D} = g_D\,|\hat{d}\cdot\vec{\Delta}|^2$,
which is minimized if $\hat{d}$ {\sl locally} orients perpendicular to the two dominant orbital components.
However, spatial variations of the order parameter cost gradient energy. In the case of the orbital components the 
spatial profiles are already optimized by minimizing the GL functional subject to the boundary conditions of the 
confining geometry. 
For the inhomogeneous phases of superfluid \He{} in pore of radius $R=100\nm{}$, the spatial variations of the 
orbital components occur on a length scale that is short compared to the dipole coherence length, 
$\xi_D\equiv \sqrt{g_D/K_1}\approx 10\,\mu$m.
Thus, spatial variations of $\hat{d}$ on such short length scales of the pore geometry cost much more than the 
dipole energy. As a result $\hat{d}$ ``unlocks'' from the {\sl local} variations of the orbital order parameter. 
This allows us to average the orbital components over the cross-section of the cylindrical pore and treat the spin 
degrees of freedom as spatially uniform on the scale of $R$.\cite{}
For the non-ESP \Bi{} phase the spin structure is described by an orthogonal matrix, ${\bf R}[\alpha,\beta,\gamma]$,
representing the {\sl relative} rotation of the spin and orbital coordinates.
 
The orientation of the spin coordinates of the Cooper pairs is also influenced by the nuclear Zeeman energy, 
\begin{equation}
\label{eq-Zeeman}
\Delta \Omega_{Z} = g_{z}\,\int_{V}d^3r\,H_{\alpha}\left(A_{\alpha i}A_{\beta i}^{*} \right)\,H_{\beta}
\,,
\end{equation}
where 
\begin{equation}
g_z = \frac{N(0)\gamma^2\hbar^2}{(1 + F_0^a)^2}\,\frac{7\zeta(3)}{48\pi^2 T_{c}^2} > 0
\,, 
\end{equation}
is the Zeeman coupling constant in the weak-coupling limit.
For ESP states the static NMR field prefers $\hat{d}\perp\vec{H}$.

\begin{figure}[t]
\begin{center}
\includegraphics[width=3.2in]{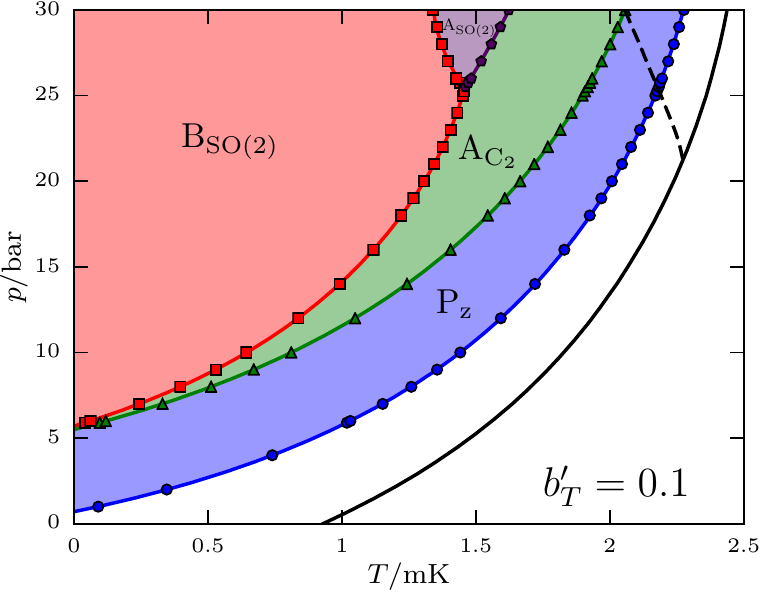}
\caption{Phase diagram for $b_T'=0.1$ using the $\beta$ parameters of Sauls and Serene.\cite{sau81b}
The resulting phase boundaries are largely left unchanged except for the shift upward in pressure of the 
tricritical point, roughly corresponding to the difference of the Choi et al\cite{cho07} polycritical 
point, $\pPCP{} \approx 21\mbar{}$, and the Sauls and Serene polycritical point, $\pPCP{} \approx 28\mbar{}$.}
\label{phase_diagram_ss81}
\end{center}
\end{figure}

\vspace*{-3mm}
\subsection{Fast vs. slow spin dynamics}
\vspace*{-3mm}

The nuclear dipolar energy generates frequency shifts, $\Delta\omega = \omega-\omega_L$,
of the NMR resonace line for superfluid \He{} away from the Larmor 
frequency, $\omega_L=\gamma H$, that are sensitive to the spin and orbital
structure of the ordered phase, the strength and orientation of the static NMR field, and the {\sl rf} 
field (``tipping field'') used to excite the nuclear spins.
In the high field limit, $\omega_{L} \gg \Omega$, where $\Omega\sim\Omega_{A}$ is the
dynamical timescale set by the dipole energy, 
we use Fomin's formulation of the spin dynamics based on the separation of fast and slow timescales 
for the dynamics of the magnetization (see also Ref. \onlinecite{bun93}), or total spin, $\vec{S}(t)$, and
the order parameter, ${A}(t)$. The ``fast'' response is on the scale set by the Larmor 
freqency, $\omega\sim\omega_L$, while the ``slow'' response is set by the dipolar frequency of 
order $\Omega_A$.\cite{fom78}
Note that the static NMR field is still assumed to be small in the sense that the Zeeman energy is 
much smaller than the condensation and gradient energies associated with the orbital components of the 
order parameter. Thus, the dynamical contributions to the nuclear dipole and Zeeman energies can be 
calculated on the basis of the solutions for the orbital order parameter in zero field.
However, for static NMR fields greater than the Dipole field, $H\gg H_{D}\approx 30\,\mbox{G}$, the 
equilibrium orientation of the spin components of the order parameter is determined primarily by the Zeeman 
energy, with the dipole energy resolving any remaining degeneracy in the equilibrium orientation of the 
$\hat{d}$ vector, or the rotation matrix ${\bf R}$ for the \Bi{} state. This provides us with the initial 
equilibrium conditions for orientation of the spin components of the order parameter.

The spin dynamics of the superfluid phases is parametrized in terms of rotation matrices for the precession 
of the order parameter, e.g. ${A}(t)$ and total spin, $\vec{S}(t)$, following an initial {\sl rf} excitation 
of the spin system. An {\sl rf} impulse applied at $t=0$ rotates (``tips'') the total spin, $\vec{S}(t=0^+)$ 
by an angle $\beta$ relative to the equilibrium spin, $\vec{S}_0 || \vec{H}\equiv H\hat{\mathbf z}'$.
The resulting dynamics of the order parameter for timescales, $0<t\ll 2\pi/\Omega$, is the 
parametrized by\cite{fom78}
\begin{equation}
\label{eq-OP-spin_dynamics}
{A}(t)=\mathbf{R_{z'}}(\omega t)\mathbf{R_{y'}}(\beta)\mathbf{R_{z'}}(-\omega t+\vartheta){A}_0 
\,,
\end{equation}
where ${\mathbf y'}\perp{\mathbf z'}$ is the direction of the {\sl rf} tipping field, and the rotation angles 
are defined by one ``fast'' angle, $\omega t$, and two ``slow'' dynamical angles, $\beta$ and $\vartheta$.
Inserting this expression into Eq. \ref{eq-dipole_energy} for the dipole energy and averaging the result over the 
fast time scale, $2\pi/\omega$, gives the fast-time and short-distance scale averaged dipole energy density,
\begin{equation}
\overline{f_{D}} = \frac{\omega}{2\pi}\int_{0}^{\frac{2\pi}{\omega}}\negthickspace 
dt\,\frac{1}{V_{\text{cell}}} \int\,d^3r\,f_{D}[{A}(\vec{r},t)]
\,.
\label{eq-average_dipole_energy}
\end{equation}
This averaged dipolar energy functional determines the transverse NMR frequency shift $\Delta\omega$ as a 
function of tipping angle $\beta$ for various orientations of the NMR field, $\vec{H}$, relative to the 
order parameter within the pore geometry.
The variable $\vartheta$ -- the generalization of Leggett's rotation angle for the bulk B-phase -- is fixed by 
the stationary condition of $\overline{f_{D}}$. The transverse NMR frequency shift as a function of tipping angle 
is then given by\cite{fom78}
\begin{equation}
\omega\Delta\omega = \frac{\gamma^2}{\chi}
		     \frac{1}{\sin{\beta}}\frac{d}{d\beta}\overline{f_{D}}
\,,
\label{eq-tranverse-general}
\end{equation}
which provides the key NMR signatures for the confined phases of \He{} under strong confinement.

\vspace*{-3mm}
\subsection{\Pz{} and \Ai{} Phases}
\vspace*{-3mm}

\begin{figure}[t]
\begin{center}
\includegraphics[width=2.6in]{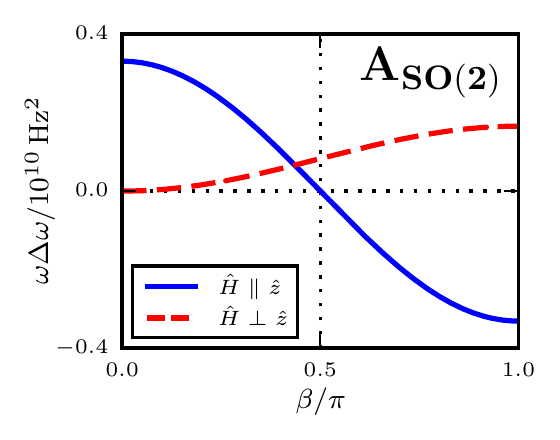}
\caption{Frequency shift of the \Ai{} phase at $p=26\,\mbox{bar}$, 
         $T=0.5\,T_c$, and $b_T'=0$. The \Pz{} phase has the same functional form, but with larger amplitude.}
\end{center}
\end{figure}

The \Pz{} and \Ai{} phases are both ESP phases parameterized by a real $\hat{d}$ vector with order 
parameters given by Eqs. \ref{eq:pz} and \ref{eq:ai}, respectively.
Spatial averaging of the dipole energy for these two phases leads to a dipole energy of the same form for both phases,
\begin{equation}
f_{D} = g_D \left(2\Dzz-\Dtt\right)\left(\hat{d}\cdot\hat{z}\right)^{2}
\,,
\label{eq-dipole_energy_Pz+ASO2}
\end{equation}
where $\hat{z}$ is the axis of the cylindrical pore, and $\hat{d}$ is homogeneous
and oriented in equilibrium in the plane perpendicular to the NMR field $\vec{H}$ and along 
a direction that minimizes Eq. \ref{eq-dipole_energy_Pz+ASO2}.

Parameterizing the direction of the NMR field in the coordinate system of the cylindrical pore by 
$\mathbf{\hat{z}'}=\left\lbrace \cos\phi\sin\theta,\,\sin\phi\sin\theta,\,\cos\theta \right\rbrace$, 
carrying out the transformation Eq. \ref{eq-OP-spin_dynamics} yields $\overline{f_D}$ in
Eq. \ref{eq-average_dipole_energy},
\begin{align}
\overline{f_D} 
&= \frac{1}{8}g_D\left(\Dtt - 2\Dzz\right)\left[2(\cos{\beta}+1)^2 \cos^2{\vartheta} \sin^2{\theta} \right. 
\nonumber\\
&\quad +\left. 4 \cos^2{\beta} - (2\cos{\beta}+7\cos^2{\beta})\sin^2{\theta} \right]\,.
\label{eq-average_dipole_Pz}
\end{align}
Since $\langle \Delta_\phi^2 \rangle - 2\langle \Delta_z^2 \rangle < 0$ in the pore, $\overline{f_D}$ is minimized 
with respect to $\vartheta$ with $\vartheta=0$. Finally, a general expression for the transverse shifts as a function 
of tipping angle is obtained with Eq. \ref{eq-tranverse-general},
\begin{align}
\omega\Delta\omega
&= 
\wpre{}
\left(2\Dzz-\Dtt\right)\nonumber\\
&\quad\times \left[\cos(\beta)-\sin^{2}(\theta)\left(\frac{5 \cos(\beta) -1}{4}\right)\right]
\,.
\label{shift-Pz}
\end{align}
The dependences on the tipping angle, $\beta$, and the polar orientation of the 
NMR field, $\theta$, are identical for both the \Pz{} and \Ai{} phases - only the magnitude of the shift differs
between the two phases. Note in particular that the shift vanishes precisely at $\beta=\pi/2$ for $\vec{H}||\hat{z}$. 
The result for the \Ai{} phase is equivalent to what is predicted for a 2D orbital glass phase of \He-A.\cite{dmi10}
Although the \Pz{} and \Ai{} phases differ only quantitatively in their transverse NMR frequency shift, 
they can still be distinguished in sufficiently clean pores by the change in temperature dependence near the 
second order phase transition between the two phases (see Fig. \ref{phase_diagram_min}), in particular the 
discontinuity in the derivative of the frequency shift, $d\Delta\omega/dT|_{T_{c_2}}$. 

\vspace*{-3mm}
\subsection{\At{} Phase}
\vspace*{-3mm}

The \At{} phase, with order parameter given by Eq. \ref{eq:at}, breaks $\SO(2)$ orbital symmetry,
and exhibits distinctly different NMR signatures compared to those of the \Ai{} phase. 
Here we consider the two cases $\vec{H} \parallel \hat{z}$ and $\vec{H} \perp \hat{z}$. 
For $\vec{H}\perp\hat{z}$ the residual $D_{2h}$ symmetry leads to a dependence of the transverse 
frequency shift on the azimuthal angle of $\vec{H}$. 
Due to the $\phi$ dependence of the order parameter it is convenient to work in Cartesian coordinates 
with the chiral axis fixed along $\langle\vec{l}\rangle\parallel\hat{y}$. The resulting 
spatial averages of the order parameter profiles entering the average dipole energy become,
\begin{align}
\Dxx &= \savg{(\Delta_r \cos\phi - \Delta_\phi \sin\phi)^2} \nonumber\\
\Dyy &= \savg{(\Delta_r \sin\phi + \Delta_\phi \cos\phi)^2} \nonumber\\
\savg{\Delta_x\Delta_y} &= \savg{(\Delta_r \cos\phi - \Delta_\phi \sin\phi)\right. \nonumber\\
 &\quad\times \left. (\Delta_r \sin\phi + \Delta_\phi \cos\phi)}
= 0 
\,,
\end{align}

\begin{figure}[t]
\begin{center}
\includegraphics[width=2.6in]{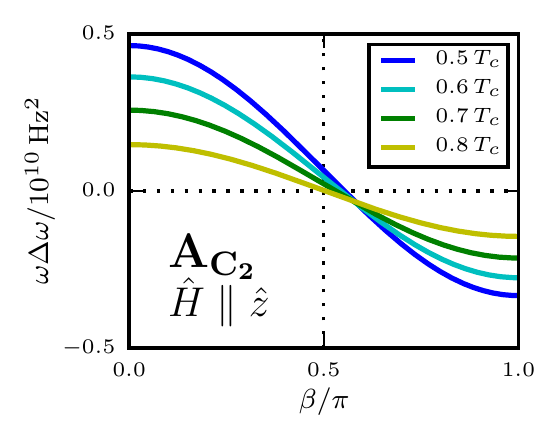}
\caption{Transverse frequency shifts for the \At{} phase with $\vec{H}\parallel\hat{z}$ at $p=26\,\mbox{bar}$,
         temperatures $T=0.5-0.8\,T_c$, and maximal pairbreaking with $b_T'=0$.}
\label{fig-NMR_Ac2_H||z}
\end{center}
\end{figure}

For $\vec{H}\parallel\hat{z}$, the result for the transverse frequency shift for the \At{} phase becomes
\begin{align}
\omega\Delta\omega^{(\parallel)}
&= \frac{1}{2}\wpre{}
\left\lbrace \Dyy-\Dxx \right. 
\nonumber \\
&\quad \left. -\left(3\Dxx+\Dyy-4\Dzz \right)\cos\beta \right. \nonumber\\
&\quad \left.+2\left|\Dyy-\Dxx \right|(1+\cos\beta)
\right\rbrace 
\,.
\end{align}
The results for $\omega\Delta\omega$ for several temperatures are shown in Fig. \ref{fig-NMR_Ac2_H||z}. 
The shift is similar to that for the \Pz{} and \Ai{} phases, except for the  asymmetry of the positive 
and negative shifts. Note also that $\Delta\omega$ vanishes at a temperature-dependent angle $\beta^* > \pi/2$.  

\begin{figure}[t]
\begin{center}
\includegraphics[width=1\linewidth]{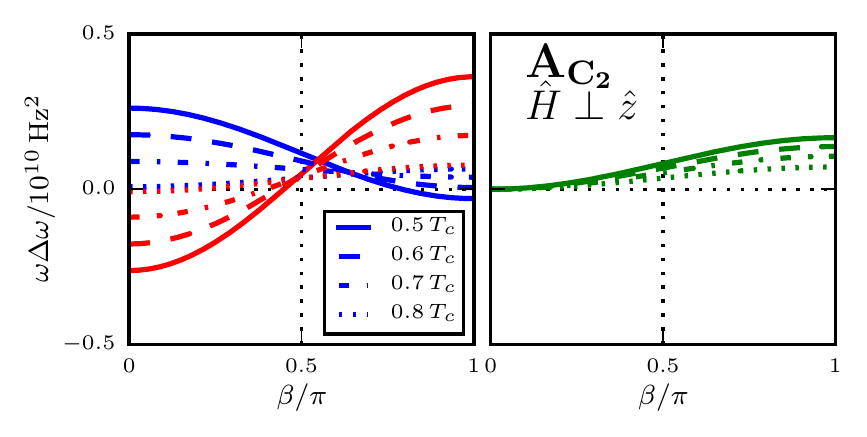}
\caption{Transverse frequency shifts for the \At{} phase with $\vec{H}\perp\hat{z}$ at $p=26\,\mbox{bar}$,
temperatures $T=0.5-0.8\,T_c$, and maximal pairbreaking with $b_T'=0$.
The left panel shows the shifts for in-plane field orientation $\varphi=0$ (blue),
and $\varphi=\pi/2$ (red). The right panel shows the shifts for $\varphi=\pi/4$, 
which has the same functional form as that of the \Ai{} phase.}
\label{fig-NMR_Ac2_H_perp_z}
\end{center}
\end{figure}

In contrast, for $\vec{H}=H(\cos\varphi\hat{x}+\sin\varphi\hat{y})$ the shift, $\omega\Delta\omega$, 
depends on the azimuthal angle $\varphi$ of the static field, in addition to the tipping angle $\beta$.
For an order parameter of the form in Eq. \ref{eq:at} we have $\langle\vec{l}\rangle\parallel\hat{y}$, and
the resulting transverse frequency shift as a function of $\varphi$ and $\beta$ becomes,
\begin{align}
\omega\Delta\omega^{(\perp)}
&= \frac{1}{4}\wpre{} 
\left\lbrace
\left(\Dxx+\Dyy-2\Dzz\right)(\cos\beta-1) 
\right. 
\nonumber \\
&\quad \left.
+\left(\Dxx-\Dyy\right)(1+7\cos\beta)\cos2\varphi
\right\rbrace 
\,.
\end{align}
The results for several field orientations, $\varphi=0,\pi/4\,,\pi/2$, are shown in Fig. \ref{fig-NMR_Ac2_H_perp_z}.
The tipping angle dependences for $\vec{H}\parallel\langle\vec{l}\rangle$ and 
$\vec{H}\perp\langle\vec{l}\rangle$ are of the same functional form as the corresponding cases for bulk \He-A.
There is a ``magic'' tipping angle of $\beta_{x}=\cos^{-1}(-1/7)\approx 0.545\pi$ at 
which $\Delta\omega(\varphi=0)=\Delta\omega(\varphi=\pi/2)$ independent of temperature.
The tipping angle dependence for $\varphi=\pi/4$ is much weaker, and qualitatively similar to that for the 
\Ai{} phase with $\vec{H}\perp\hat{z}$.
Observation of these results for several field orientations would provide a clear identification of the 
\At{} phase and determine the direction of the mean chiral axis.

\vspace*{-3mm}
\subsection{\Bi{} Phase}
\vspace*{-3mm}

The \Bi{} phase is a non-ESP phase with a reduced and anisotropic spin susceptibility below $T_c$. 
The \Bi{} phase also exhibits tipping angle dependence of the frequency shift similar to that of \He-B. 
In particular, for $\vec{H}\perp \hat{z}$ the shift is a polar-distorted Brinkman-Smith mode, and for  
$\vec{H}\parallel\hat{z}$ we find the ``perpendicular'' mode that is qualitatively similar, but 
with important quantitative differences compared to bulk \He-B.

\begin{figure}[t]
\begin{center}
\includegraphics[width=1\linewidth]{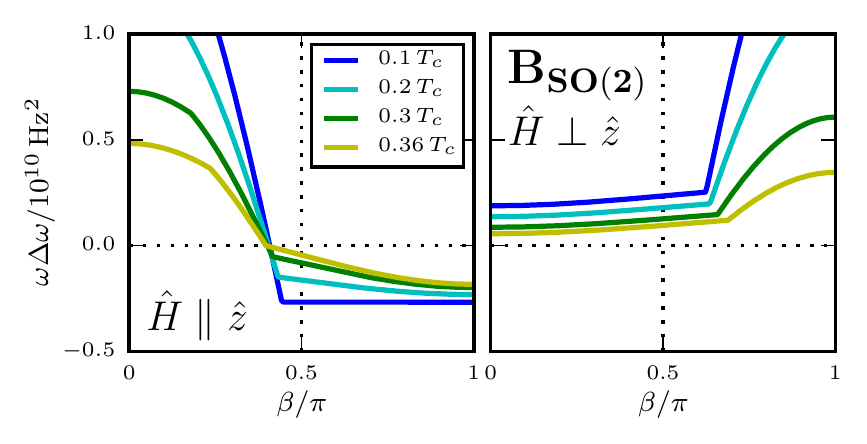}
\caption{Transverse frequency shifts for the \Bi{} phase at $p=10\,\mbox{bar}$, temperatures 
$T=0.1\,T_c$, 
$0.2\,T_c$, 
$0.3\,T_c$, 
$0.36\,T_c$, 
and maximal pair-breaking, $b_T'=0$.}
\label{nmr_bi_par}
\end{center}
\end{figure}

For $\vec{H} \parallel\hat{z}$, the Zeeman energy is minimized in equilibrium by a spin-rotation  
$\mathbf{R_{\hat{n}}}(\pi / 2)$, where $\hat{n}$ is in the $x-y$ plane.
This rotation leads to a positive transverse frequency shift that is maximal at small tipping angles, 
unlike the Brinkman-Smith mode. The quantitative description of the frequency shift depends on the 
spatial averages of the \Bi{} gap parameters,
\begin{align}
P &= \Drz+\Dtz \nonumber\\
Q &= \Drr+2\Drt+\Dtt\ \nonumber\\
R &= \Drr+\Dtt+4\Dzz 
\,,
\end{align}
with the resulting transverse shift for $\vec{H}\parallel\hat{z}$ given by
\begin{align}\label{eq-omegaBparallel}
\omega\Delta\omega^{(\parallel)}
&= \frac{1}{8}\wpreB{}
\left\lbrace
 4R\,\cos\beta
+4P\,(1+4\cos\beta)\cos\vartheta 
\right.
\nonumber \\
&\quad 
\left.
\qquad
\qquad
-Q\,(1+\cos\beta)\cos2\vartheta
\right\rbrace 
\,,
\end{align}
where the Leggett angle is
\begin{align}
\cos\vartheta 
&= \left\{
\begin{array}{ll} +1 & : \beta \le \beta_L' 
\,,
\\
\frac{2P(2\cos\beta-1)}{Q(1+\cos\beta)} & : \beta_L' < \beta < \beta_L
\,,
\\
-1 & : \beta \geq \beta_L
\,,
\end{array}
\right.
\end{align}
with $\cos\beta_L'=(2P+Q)/(4P-Q)$ and $\cos\beta_L \equiv (2P-Q)/(4P+Q)$.\cite{Note2}
The left panel of Fig. \ref{nmr_bi_par} shows the tipping-angle dependence 
of the frequency shift $\Delta\omega^{(||)}$ for temperatures starting just below the transition to \Bi{}-\At{} 
phase boundary at $p=10\,\text{bar}$. At this pressure transition from the \Pz{} phase to the \Bi{} phase is 
interrupted by a narrow sliver of \At{} phase. Thus, near the \Bi{}-\At{} phase boundary, the \Pz{} order parameter 
is dominant and that is reflected in the tipping angle dependence for $T=0.36T_c$. At lower temperatures the 
transverse components of the \Bi{} become significant and the transverse shift evovles towards a form characteristic 
of the B-phase with a sharp transition at $\beta_L$. The polar distortion is still manifest as the negative shift 
for $\beta>\beta_L$.  

For the static NMR field $\vec{H}\perp\hat{z}$ the relevant averages of the \Bi{} gap are
\begin{align}
P &= \Drr+\Dtt+2\Drt+2\Drz+2\Dtz \nonumber\\
Q &= 3\Drr+3\Dtt+2\Drt \nonumber\\
&\quad+8\Drz+8\Dtz+8\Dzz \nonumber\\
R &= 11\Drr+11\Dtt+18\Drt+8\Dzz
\,,
\end{align}
leading to a transverse shift
\begin{align}\label{eq-omegaBperp}
\omega\Delta\omega^{(\perp)}
&= -\frac{1}{16}\wpreB{} 
\left\lbrace
2R\,\cos\beta
+4P\,(1+4\cos\beta)\cos\vartheta 
\right. 
\nonumber\\
&\quad \left. +Q\,(1+\cos\beta)\cos2\vartheta
\right\rbrace 
\,,
\end{align}
where
\begin{align}
\cos\vartheta &= \left\{\begin{array}{lr}
\frac{-2P(2\cos\beta-1)}{Q(1+\cos\beta)} & : \beta < \beta_L. \\
1 & : \beta \geq \beta_L.
\end{array}\right.
\end{align}
and $\cos\beta_L = (2P-Q)/(4P+Q)$.
Unlike bulk \He-B, the confinement induced anisotropy of the 
\Bi{} order parameter results in a nonzero transverse frequency shift even at small tipping angles,
and a temperature dependent critial angle $\beta_L$. Once again the frequency shift shows the evolution
from a functional form close to the \Pz{} phase for $T=0.36T_c$ towards the polar distored B-phase
at low temperature.

\vspace*{-3mm}
\section{Summary and Outlook}
\vspace*{-3mm}

For \He{} in a long cylindrical pore of radius $R=100\nm{}$, the relative stability of superfluid phases is 
strongly dependent on pressure both through the combination of strong-coupling corrections to the fourth-order 
GL free energy and changes in the effective confinement ratio $R/\xi_0(p)$, and the degree of pairbreaking by
boundary scattering. We find four different equilibrium phases over the full pressure range for boundary
conditions spanning the range from maximal pairbreaking (retro-reflective boundaries) to minimal pairbreaking
(specular reflective boundaries). The first instability is to the $z$-aligned polar \Pz{} phase, except for the 
idealized case of perfect specular reflection for a perfectly circular pore. A polar distorted B-like phase
is stabilized at sufficiently low temperatures within our theory for the strong-coupling effects based on 
the GL functional. We find two symmetry inequivalent chiral A-like phases, the axially symmetric 
\Ai{} phase with a radially directed chiral field and a polar core favored in the limit of weak pairbreaking,
and the broken axial symmetry chiral \At{} phase with chirality directed perpendicular to the axis of the pore.
The \Ai{} phase dominates the phase diagram for specularly reflecting boundaries while the \At{} phase appears
at intermediate temperatures and higher pressures separating the \Pz{} and \Bi{} phases. 
The four equilibrium phases can be identified by their distinct NMR freqency shifts as functions of tipping angle 
and NMR field orientation. NMR experiments utilizing arrays of equivalent nano-pores should be able to test these 
predictions and uniquely identify the polar phase as well as the new prediction of the broken symmetry 
chiral \At{} phase. 

The interplay of complex symmetry breaking, spatial confinement, surface disorder and strong-interactions beyond 
weak-coupling BCS leads to a remarkably rich phase diagram of broken symmetry states in what is perhaps
the simplest of confining geometries, the cylindrical pore.
We expect an even wider spectrum of broken symmetry phases with unique physical properties in more complex confining 
geometries,\cite{wim13} or when confinement is in competition with external fields or the formation of topological 
defects.\cite{vor05,vor07} Indeed theoretical reports of new phases of superfluid \He\ in thin films and cavities have 
simulated the development of nano-scale cavities, MEMS and nano-fluidic oscillators and new nano-scale materials for 
experimental search and discovery of new quantum ground states.
In the latter category the infusion of quantum fluids into highly porous anisotropic aerogels has opened a new window
into the role of confinement on complex symmetry breaking. New chiral and ESP phases of superfluid \He\ in uniaxially 
stretched and compressed silica aerogels have been reported,\cite{ben11,pol12,sau13} and in a new class of nano-scale 
confining media, called ``nematic'' aerogels, there is strong evidence to support the observation of a polar \Pz{} phase 
of \He\ in this strongly anisotropic {\sl random} medium.\cite{ask12,dmi15c}
From the vantage point of our predictions for \He\ confined in a long cylindrical pore there are strong similarities between
the phase diagram for $R=100\,\mbox{nm}$ cylindrical pores and the experimental phase diagram of \He\ in nematic aerogels, 
including the normal to \Pz{} transition, and uniaxially deformed B-like and chiral A-like phases. It is remarkable
that the subtle correlations giving rise to chirality of an \At{} or \Ai{} phase survives the random potential 
of these disordered porous solids. The observations pose challenges for theorists to provide a quantitative understanding 
of how complex symmetry breaking and long-range order remain so robust in random anisotropic materials.


\vspace*{-5mm}
\section{Material Parameters}\label{sec-Appendix} 
\vspace*{-3mm}

The following tables summarize the pressure dependent material parameters that determine the superfluid 
phases in strong-coupling GL theory.
 
\vspace*{-3mm}
\begin{table}[H]
\begin{center}\tiny
\begin{tabular}{|c|c|c|c|c|c|c|c|c|}
\hline
$p$[bar] & $T_c$[mK] & $v_f$[m/s] & $\xi_0$[nm] & $\scb{1}$ & $\scb{2}$ & $\scb{3}$ & $\scb{4}$ & $\scb{5}$\cr
\hline
    0.0   &  0.929 &  59.03 & 77.21 & 0.03 & -0.11 & 0.10 & -0.15 & 0.16\cr
    2.0   &  1.181 &  55.41 & 57.04 & 0.03 & -0.04 & -0.14 & -0.37 & 0.19\cr
    4.0   &  1.388 &  52.36 & 45.85 & 0.02 & -0.01 & -0.24 & -0.48 & 0.19\cr
    6.0   &  1.560 &  49.77 & 38.77 & 0.02 & -0.01 & -0.28 & -0.54 & 0.18\cr
    8.0   &  1.705 &  47.56 & 33.91 & 0.02 & -0.02 & -0.30 & -0.58 & 0.17\cr
   10.0   &  1.828 &  45.66 & 30.37 & 0.01 & -0.03 & -0.31 & -0.60 & 0.15\cr
   12.0   &  1.934 &  44.00 & 27.66 & 0.01 & -0.04 & -0.31 & -0.61 & 0.13\cr
   14.0   &  2.026 &  42.51 & 25.51 & 0.00 & -0.05 & -0.30 & -0.62 & 0.11\cr
   16.0   &  2.106 &  41.17 & 23.76 & 0.00 & -0.05 & -0.27 & -0.66 & 0.10\cr
   18.0   &  2.177 &  39.92 & 22.29 & 0.00 & -0.06 & -0.27 & -0.68 & 0.09\cr
   20.0   &  2.239 &  38.74 & 21.03 & -0.01 & -0.06 & -0.26 & -0.69 & 0.07\cr
   22.0   &  2.293 &  37.61 & 19.94 & -0.01 & -0.07 & -0.26 & -0.71 & 0.06\cr
   24.0   &  2.339 &  36.53 & 18.99 & -0.01 & -0.07 & -0.26 & -0.72 & 0.04\cr
   26.0   &  2.378 &  35.50 & 18.15 & -0.02 & -0.07 & -0.27 & -0.73 & 0.03\cr
   28.0   &  2.411 &  34.53 & 17.41 & -0.02 & -0.07 & -0.27 & -0.74 & 0.01\cr
   30.0   &  2.438 &  33.63 & 16.77 & -0.02 & -0.07 & -0.28 & -0.74 & -0.01\cr
   32.0   &  2.463 &  32.85 & 16.22 & -0.03 & -0.07 & -0.27 & -0.75 & -0.02\cr
   34.0   &  2.486 &  32.23 & 15.76 & -0.03 & -0.07 & -0.27 & -0.75 & -0.03\cr
\hline
\end{tabular}
\caption{Material parameters for \He{} vs. pressure, with $T_c$ from Ref. \onlinecite{gre86}, $v_f$ 
calculated with $m^{*}$ from Ref. \onlinecite{gre86} and density $n$ from Ref. \onlinecite{whe75}. 
Coherence lengths are calculated as $\xi_0=\hbar v_f/2\pi\,k_B T_c$. Strong-coupling $\scb{i}$ parameters 
at $T_c$ in units of $|\beta_1^{\text{wc}}|$ are from Ref. \onlinecite{cho07}.}
\label{table:3He-Material-Parameters}
\end{center}
\end{table}
\vspace*{-3mm}
\begin{table}[H]
\begin{center}\tiny
\begin{tabular}{|c|r|r|r|r|r|}
\hline
\multicolumn{1}{|c|}{$n$} & 
\multicolumn{1}{|c|}{$\scb{1}$} & 
\multicolumn{1}{|c|}{$\scb{2}$} & 
\multicolumn{1}{|c|}{$\scb{3}$} & 
\multicolumn{1}{|c|}{$\scb{4}$} & 
\multicolumn{1}{|c|}{$\scb{5}$}\cr
\hline
0   &  \sn{3.070}{-2}   &  \sn{-1.074}{-1}  & \sn{1.038}{-1} 	& \sn{-1.593}{-1} 	& \sn{1.610}{-1}\cr
1   &  \sn{-2.081}{-3}  &  \sn{5.412}{-2}   & \sn{-1.752}{-1} 	& \sn{-1.350}{-1} 	& \sn{2.263}{-2}\cr
2   &  \sn{2.133}{-5}   &  \sn{-1.081}{-2}  & \sn{3.488}{-2} 	& \sn{1.815}{-2} 	& \sn{-4.921}{-3}\cr
3   &  \sn{-4.189}{-7}  &  \sn{1.025}{-3}   & \sn{-4.243}{-3} 	& \sn{-1.339}{-3} 	& \sn{3.810}{-4}\cr
4   &---\hphantom{77777}&  \sn{-5.526}{-5}  & \sn{3.316}{-4} 	& \sn{5.316}{-5} 	& \sn{-1.529}{-5}\cr
5   &---\hphantom{77777}&  \sn{1.722}{-6}   & \sn{-1.623}{-5} 	& \sn{-1.073}{-6} 	& \sn{3.071}{-7}\cr
6   &---\hphantom{77777}&  \sn{-2.876}{-8}  & \sn{4.755}{-7} 	& \sn{8.636}{-9} 	& \sn{-2.438}{-9}\cr
7   &---\hphantom{77777}&  \sn{1.991}{-10}  & \sn{-7.587}{-9} 	&---\hphantom{77777}  	&---\hphantom{77777} \cr
8   &---\hphantom{77777}&---\hphantom{77777}& \sn{5.063}{-11} 	&---\hphantom{77777}  	&---\hphantom{77777} \cr
\hline
\end{tabular}
\caption{Coefficients of a polynomial fit to the strong-coupling $\beta$ parameters from Ref. \onlinecite{cho07} 
         of the form $\scb{i} = \sum_n a^{(i)}_n \, p^n$.}
\label{table:3He-Material-Parameters_Fit-Choi}
\end{center}
\end{table}
\vspace*{-3mm}
\begin{table}[H]
\begin{center}\tiny
\begin{tabular}{|c|c|c|c|c|c|}
\hline
$p$[bar] & $\scb{1}$ & $\scb{2}$ & $\scb{3}$ & $\scb{4}$ & $\scb{5}$\cr
\hline
    0    &  -0.008 &  -0.033 & -0.043 & -0.054 & -0.055\cr
    12   &  -0.034 &  -0.080 & -0.117 & -0.199 & -0.194\cr
    16   &  -0.041 &  -0.088 & -0.129 & -0.230 & -0.236\cr
    20   &  -0.048 &  -0.095 & -0.136 & -0.254 & -0.277\cr
    24   &  -0.055 &  -0.101 & -0.140 & -0.272 & -0.320\cr
    26   &  -0.059 &  -0.103 & -0.140 & -0.280 & -0.344\cr
    28   &  -0.062 &  -0.105 & -0.139 & -0.287 & -0.370\cr
    30   &  -0.066 &  -0.106 & -0.137 & -0.292 & -0.398\cr
    32   &  -0.070 &  -0.106 & -0.132 & -0.296 & -0.429\cr
    34.4 &  -0.074 &  -0.103 & -0.123 & -0.298 & -0.469\cr
\hline
\end{tabular}
\caption{Sauls-Serene $\Delta\beta_{i}^{\text{sc}}$ parameters\cite{sau81b} for \He{} vs. pressure. The values 
         at $p=0\mbar{}$ were obtained by extrapolating the published $\Delta\beta_{i}^{\text{sc}}$, which were 
         calculated only down to $12 \mbar{}$,
         to their weak-coupling values at $\lim_{p\rightarrow p_0}T_c(p)/T_F(p)=0$, 
         which corresponds to a negative pressure of $p_0=-5\,\mbox{bar}$.}
\label{table:3He-Material-Parameters-SS81}
\end{center}
\end{table}
\vspace*{-3mm}
\begin{table}[H]
\begin{center}\tiny
\begin{tabular}{|c|r|r|r|r|r|}
\hline
\multicolumn{1}{|c|}{$n$} 
& \multicolumn{1}{|c|}{$\scb{1}$} 
& \multicolumn{1}{|c|}{$\scb{2}$} 
& \multicolumn{1}{|c|}{$\scb{3}$} 
& \multicolumn{1}{|c|}{$\scb{4}$} 
& \multicolumn{1}{|c|}{$\scb{5}$}\cr
\hline
0   &  \sn{-8.311}{-3}  &  \sn{-3.334}{-2}  & \sn{-4.298}{-2} 	& \sn{-5.416}{-2} & \sn{-5.505}{-2}\cr
1   &  \sn{-2.404}{-3}  &  \sn{-4.716}{-3}  & \sn{-7.988}{-3} 	& \sn{-1.550}{-2} & \sn{-1.427}{-2}\cr
2   &  \sn{2.813}{-5}   &  \sn{8.032}{-5}   & \sn{1.637}{-4} 	& \sn{3.174}{-4}  & \sn{2.942}{-4}\cr
3   &  \sn{-4.024}{-7}  &  \sn{-9.400}{-8}  & \sn{-1.345}{-8} 	& \sn{-2.138}{-6} & \sn{-6.654}{-6}\cr
\hline
\end{tabular}
\caption{Coefficients of a polynomial fit to the Sauls-Serene $\beta$ parameters in 
         Table \ref{table:3He-Material-Parameters-SS81} of the form $\scb{i} = \sum_n a^{(i)}_n \, p^n$.}
\label{table:3He-Material-Parameters_Fit-SS81}
\end{center}
\end{table}

\vspace*{-5mm}
\section{Acknowledgements}
\vspace*{-3mm}

The research of JJW and JAS was supported by the National Science Foundation (Grant DMR-1106315).
We acknowledge key discussions with Erkki Thuneberg and Anton Vorontsov on boundary conditions for 
the GL theory of \He{} in confined geometries, and with Bill Halperin and Andrew Zimmerman on experimental 
realizations of superfluidity with strong confinement that provided important motivation for this study.



\begin{thebibliography}{32}
\providecommand{\natexlab}[1]{#1}
\providecommand{\url}[1]{\texttt{#1}}
\expandafter\ifx\csname urlstyle\endcsname\relax
  \providecommand{\doi}[1]{doi: #1}\else
  \providecommand{\doi}{doi: \begingroup \urlstyle{rm}\Url}\fi
\footnotesize

\bibitem[Ambegaokar et~al.(1975)Ambegaokar, {de Gennes}, and Rainer]{amb75}
V.~Ambegaokar, P.~G. {de Gennes}, and D.~Rainer.
\newblock Landau-ginzburg equations for an anisotropic superfluid.
\newblock \emph{Phys. Rev. A}, 9:\penalty0 2676, 1975.

\bibitem[Levitin et~al.(2013)Levitin, Bennett, Casey, Cowan, Saunders, Drung,
  Schurig, and Parpia]{lev13}
L.~V. Levitin, R.~G. Bennett, A.~Casey, B.~Cowan, J.~Saunders, D.~Drung, Th.~Schurig,
  and J.~M. Parpia.
\newblock {Phase Diagram of the Topological Superfluid $^3$He Confined in a
  Nanoscale Slab Geometry}.
\newblock \emph{Science}, 340\penalty0 (6134):\penalty0 841--844, 2013.

\bibitem[Pollanen et~al.(2012)Pollanen, Li, Collett, Gannon, Halperin, and Sauls]{pol12}
J.~Pollanen, J.~I.~A. Li, C.~A. Collett, W.~J. Gannon, W.~P. Halperin, and J.~A. Sauls.
\newblock {New chiral phases of superfluid $^3$He stabilized by anisotropic silica aerogel}.
\newblock \emph{Nature Physics}, 8\penalty0 (4):\penalty0 317--320, 2012.

\bibitem[Askhadullin et~al.(2012)Askhadullin, Dmitriev, Krasnikhin, Martynov,
  Osipov, Senin, and Yudin]{ask12}
R.~Sh. Askhadullin, V.~V. Dmitriev, D.~A. Krasnikhin, P.~N. Martynov, A.~A. Osipov,
  A.~A. Senin, and A.~N. Yudin.
\newblock {Phase diagram of superfluid $^{3}$He in “nematically ordered”
  aerogel}.
\newblock \emph{JETP Lett.}, 95\penalty0 (6):\penalty0 326--331, 2012.

\bibitem[Chung and Zhang(2009)]{chu09}
S.~B. Chung and S.-C. Zhang.
\newblock {Detecting the Majorana Fermion Surface State of $^3$He-B through
  Spin Relaxation}.
\newblock \emph{Phys. Rev. Lett.}, 103\penalty0 (23):\penalty0 235301, 2009.

\bibitem[Nomura et~al.(2014)Nomura, Murakawa, Wasai, Akiyama, Nakao, and
  Okuda]{nom14}
R.~Nomura, S.~Murakawa, M.~Wasai, K.~Akiyama, T.~Nakao, and Y.~Okuda.
\newblock {Surface Majorana cone of the topological superfluid $^3$He-B phase}.
\newblock \emph{Physica E}, 55:\penalty0 42--47, 2014.

\bibitem[Mizushima and Machida(2011)]{miz11}
T.~Mizushima and K.~Machida.
\newblock {Effects of Majorana Edge Fermions on Dynamical Spin Susceptibility
  in Topological Superfluid $^3$He-B}.
\newblock \emph{J. Low Temp. Phys.}, 162\penalty0 (3-4):\penalty0 204--211,
  2011.

\bibitem[Mizushima(2012)]{miz12}
T.~Mizushima.
\newblock {Superfluid $^3$He in a restricted geometry with a perpendicular
  magnetic field}.
\newblock \emph{Phys. Rev. B}, 86\penalty0 (9):\penalty0 094518, 2012.

\bibitem[Wu and Sauls(2013)]{wu13}
H.~Wu and J.~A. Sauls.
\newblock {Majorana excitations, spin and mass currents on the surface of
  topological superfluid $^3$He-B}.
\newblock \emph{Phys. Rev. B}, 88\penalty0 (18):\penalty0 184506, 2013.

\bibitem[Maki(1978)]{mak78}
K.~Maki.
\newblock {Planar textures in superfluid $^3$He-A}.
\newblock \emph{J. Low Temp. Phys.}, 32\penalty0 (1-2):\penalty0 1--17, 1978.

\bibitem[Fetter and Ullah(1988)]{fet88}
A.~L. Fetter and S.~Ullah.
\newblock {Superfluid density and critical current of $^3$He in confined
  geometries}.
\newblock \emph{J. Low Temp. Phys.}, 70\penalty0 (5-6):\penalty0 515--535,
  1988.

\bibitem[Takagi(1987)]{tak87}
T.~Takagi.
\newblock {AB Phase Transition of Superfluid $^3$He Confined in Pore Geometry}.
\newblock \emph{Prog. Theor. Phys.}, 78\penalty0 (3):\penalty0 562--572, 1987.

\bibitem[Li and Ho(1988)]{li88}
Y.-H. Li and T.-L. Ho.
\newblock {Superfluid $^3$He in very confined regular geometries}.
\newblock \emph{Phys. Rev. B}, 38\penalty0 (4):\penalty0 2362, 1988.

\bibitem[Gould and Lee(1978)]{gou78}
C.~M. Gould and D.~M. Lee.
\newblock {Superfluid $^3$He in Narrow Cylinders}.
\newblock \emph{Phys. Rev. Lett.}, 41\penalty0 (14):\penalty0 967, 1978.

\bibitem[Saunders et~al.(1978)Saunders, Betts, Brewer, Swithenby, and
  Truscott]{sau78}
J.~Saunders, D.~S. Betts, D.~F. Brewer, S.~J. Swithenby, and W.~S. Truscott.
\newblock {Observations on Superfluid $^3$He-A in Small Cylinders: Evidence for
  a Textural Transition}.
\newblock \emph{Phys. Rev. Lett.}, 40\penalty0 (19):\penalty0 1278, 1978.

\bibitem[Bruinsma and Maki(1979)]{bru79}
R.~Bruinsma and K.~Maki.
\newblock {Textures in narrow cylinders in superfluid $^3$He-A}.
\newblock \emph{J. Low Temp. Phys.}, 37\penalty0 (5-6):\penalty0 607--625,
  1979.

\bibitem[Masuda and Fukuda(1995)]{mas95}
H.~Masuda and K.~Fukuda.
\newblock Ordered metal nanohole arrays made by a two-step replication of
  honeycomb structures of anodic alumina.
\newblock \emph{Science}, 268\penalty0 (5216):\penalty0 1466--1468, 1995.

\bibitem[Gonzalez et~al.(2013)Gonzalez, Zheng, Garcell, Lee, and Chan]{gon13}
M.~Gonzalez, P.~Zheng, E.~Garcell, Y.~Lee, and H.~B. Chan.
\newblock Comb-drive micro-electro-mechanical systems oscillators for low
  temperature experiments.
\newblock \emph{Review of Scientific Instruments}, 84\penalty0 (2):\penalty0
  025003, 2013.

\bibitem[Zhelev et~al.(2014)Zhelev, Bennett, Smith, Pollanen, Halperin, and
  Parpia]{zhe14}
N.~Zhelev, R.~G. Bennett, E.~N. Smith, J.~Pollanen, W.~P. Halperin, and J.~M. Parpia.
\newblock {Dissipation signatures of the normal and superfluid phases in
  torsion pendulum experiments with $^3$He in aerogel}.
\newblock \emph{Phys. Rev. B}, 89\penalty0 (9):\penalty0 094513, 2014.

\bibitem[Rojas and Davis(2015)]{roj15}
X.~Rojas and J.~P. Davis.
\newblock Superfluid nanomechanical resonator for quantum nanofluidics.
\newblock \emph{Phys. Rev. B}, 91\penalty0 (2):\penalty0 024503, 2015.

\bibitem[Buchholtz and Fetter(1977)]{buc77}
L.~J. Buchholtz and A.~L. Fetter.
\newblock {Textures in superfluid $^{3}$He-A: Hydrodynamic and magnetic effects
  in a cylindrical pore}.
\newblock \emph{Phys. Rev. B}, 15\penalty0 (11):\penalty0 5225, 1977.

\bibitem[Rainer and Serene(1976)]{rai76}
D.~Rainer and J.~W. Serene.
\newblock {Free energy of superfluid $^3$He}.
\newblock \emph{Phys. Rev. B}, 13\penalty0 (11):\penalty0 4745, 1976.

\bibitem[Sauls and Serene(1981)]{sau81b}
J.~A. Sauls and J.~W. Serene.
\newblock {Potential Scattering Models for the Quasiparticle Interactions in
  Liquid $^3$He}.
\newblock \emph{Phys. Rev.}, B24:\penalty0 183, 1981.

\bibitem[Choi et~al.(2007)Choi, Davis, Pollanen, Haard, and Halperin]{cho07}
H.~Choi, J.~P. Davis, J.~Pollanen, T.~M. Haard, and W.~P. Halperin.
\newblock {Strong coupling corrections to the Ginzburg-Landau theory of
  superfluid $^3$He}.
\newblock \emph{Phys. Rev. B}, 75\penalty0 (17):\penalty0 174503, 2007.

\bibitem[Greywall(1986)]{gre86}
D.~S. Greywall.
\newblock {$^3$He specific heat and thermometry at millikelvin temperatures}.
\newblock \emph{Phys. Rev. B}, 33\penalty0 (11):\penalty0 7520, 1986.

\bibitem[Wheatley(1975)]{whe75}
J.~C. Wheatley.
\newblock {Experimental properties of superfluid $^3$He}.
\newblock \emph{Rev. Mod. Phys.}, 47\penalty0 (2):\penalty0 415, 1975.

\bibitem[Sauls(2011)]{sau11}
J.~A. Sauls.
\newblock {Surface states, Edge Currents, and the Angular Momentum of Chiral
  $p$-wave Superfluids}.
\newblock \emph{Phys. Rev. B}, 84:\penalty0 214509, 2011.

\bibitem[Note1()]{Note1}
\newblock Recently Aoyama has shown that a translational symmetry breaking
  ``stripe'' B-like phase along the axis of the channel may be possible in
  cylindrical geometries, stabilized with an anisotropic boundary condition
  that implements specular reflection for scattering along $z$, but
  retro-reflection in the $r$-$\phi $ plane. This enhances $A_{\alpha z}$ on
  the boundary relative to $A_{\alpha \phi }$.\cite {aoy14} Our boundary
  condition has the opposite anisotropy. Thus, with our formulation it is
  unclear if conditions allow for an energetically stable B-like stripe phase.
  This question will be addressed in a separate report.

\bibitem[Bunkov(1993)]{bun93}
Yu.~M. Bunkov and G.~E. Volovik.
\newblock {On the Possibility of the Homogeneously Precessing Domain in Bulk $^3$He-A}.
\newblock \emph{Eur. Phys. Lett.}, 21\penalty0 (8):\penalty0 837--844,
1993.

\bibitem[Fomin(1978)]{fom78}
I.~A. Fomin.
\newblock {Solution of spin dynamics equations for $^3$He superfluid phases in a strong magnetic field}.
\newblock \emph{J. Low Temp. Phys.}, 31\penalty0 (3-4):\penalty0 509--526,
1978.

\bibitem[Dmitriev et~al.(2010)Dmitriev, Krasnikhin, Mulders, Senin, Volovik, and Yudin]{dmi10}
V.~V. Dmitriev, D.~A. Krasnikhin, N.~Mulders, A.~A. Senin, G.~E. Volovik, and A.~N. Yudin.
\newblock {Orbital glass and spin glass states of $^3$He-A in aerogel}.
\newblock \emph{Sov. Phys. JETP Lett.}, 91\penalty0 (11):\penalty0 599--606,
2010.

\bibitem[Note2()]{Note2}
\newblock Note that $\chi _B$ entering both Eqs. \ref {eq-omegaBparallel} and
  \ref {eq-omegaBperp} is given by $\chi _B = \chi _N / [1+ g_z /\chi _N
  {(\delimiter "426830A \Delta _r^2 \delimiter "526930B + \delimiter "426830A
  \Delta _\phi ^2 \delimiter "526930B )}]< \chi _N$. This result is obtained in
  both cases by minimizing the Zeeman energy for the specific field orientation.

\bibitem[Aoyama(2014)]{aoy14}
K.~Aoyama.
\newblock {Stripe order in superfluid $^3$He confined in narrow cylinders}.
\newblock \emph{Phys. Rev. B}, 89\penalty0 (14):\penalty0 140502, 
2014.

\bibitem[Wiman and Sauls(2013)]{wim13}
J.~J. Wiman and J.~A. Sauls.
\newblock {Superfluid phases of $^3$He in a periodic confined geometry}.
\newblock \emph{J. Low Temp. Phys.}, 174:\penalty0 1--14, 
2013.

\bibitem[Vorontsov and Sauls(2005)]{vor05}
A.~B. Vorontsov and J.~A. Sauls.
\newblock {Domain Walls in Superfluid $^3$He-B}.
\newblock \emph{J. Low Temp. Phys.}, 138:\penalty0 283--288, 
2005.

\bibitem[Vorontsov and Sauls(2007)]{vor07}
A.~B. Vorontsov and J.~A. Sauls.
\newblock {Crystalline Order in Superfluid {$^3$He} Films}.
\newblock \emph{Phys. Rev. Lett.}, 98\penalty0 (4):\penalty0 045301, 
2007.

\bibitem[Bennett et~al.(2011)Bennett, Zhelev, Smith, Pollanen, Halperin, and Parpia]{ben11}
R.~Bennett, N.~Zhelev, E.~Smith, J.~Pollanen, W.~Halperin, and J.~Parpia.
\newblock {Modification of the $^3$He Phase Diagram by Anisotropic Disorder}.
\newblock \emph{Phys. Rev. Lett.}, 107:\penalty0 235504, 
2011.

\bibitem[Sauls(2013)]{sau13}
J.~A. Sauls.
\newblock {Chiral phases of superfluid $^3$He in an anisotropic medium}.
\newblock \emph{Phys. Rev. B}, 88:\penalty0 214503, 
2013.

\bibitem[Dmitriev et~al.(2015)Dmitriev, Senin, Soldatov, and Yudin]{dmi15c}
V.~V. Dmitriev, A.~A. Senin, A.~A. Soldatov, and A.~N. Yudin.
\newblock {Polar phase of superfluid \He\ in anisotropic aerogel}.
\newblock \emph{arXiv}, 1507.04275:\penalty0 1--5, 
2015.

\end{thebibliography}
\end{document}